\documentclass[12pt]{article}

\usepackage{amsmath}
\usepackage{graphicx,psfrag,epsf,epstopdf}
\usepackage{enumerate}
\usepackage{natbib}
\usepackage{url} 

\usepackage{amssymb}
\usepackage{amsthm}
\usepackage{color}
\usepackage{caption}
\usepackage{booktabs}
\usepackage{bbm}
\usepackage{multicol}
\usepackage{multirow}
\usepackage{rotating}
\usepackage{pdflscape}

\def\bmath0{{\boldmath 0}}
\def\bm1{{\boldsymbol 1}}
\def\bmb{{\boldsymbol b}}

\def\bmx{{\boldsymbol x}}
\def\bmy{{\boldsymbol y}}

\def\bmv{{\boldsymbol v}}
\def\bmp{{\boldsymbol p}}
\def\bmr{{\boldsymbol r}}
\def\bmz{{\boldsymbol z}}
\def\bmA{{\boldsymbol A}}

\def\bmX{{\boldsymbol X}}
\def\bmY{{\boldsymbol Y}}
\def\bmR{{\boldsymbol R}}
\def\bmt{{\boldsymbol t}}
\def\bmm{{\boldsymbol \mu}}
\def\bmU{{\boldsymbol U}}
\def\bmZ{{\boldsymbol Z}}

\def\ave{{\mbox{ave}}}
\def\conv{{\mbox{conv}}}
\def\interior{{\mbox{int}}}
\def\vol{\mbox{vol}}
\def\med{\mbox{med}}
\def\mad{\mbox{MAD}}

\def\IR{{\mathbb{R}}}
\def\IL{{\mathbb{L}}}

\def\LBL{\textsf{\textbf{---}}} 
\def\LBD{{\textbf{- -}}} 

   \def\0{\boldsymbol 0}

\begin{document}

\title{\LARGE\bf Choosing among notions of multivariate depth statistics}
\author{    Karl Mosler \\    {\small Institute of Econometrics and Statistics, University of Cologne}\\
    Pavlo Mozharovskyi \\    {\small LTCI, T{\'e}l{\'e}com Paris, Institut Polytechnique de Paris}}
    
\date{May 5, 2021}
\maketitle

\vspace{5ex}

\begin{abstract}
Classical multivariate statistics measures the outlyingness of a point by its Mahalanobis distance from the mean, which is based on the mean and the covariance matrix of the data.
A {multivariate} depth function is a function which, given a point and a distribution in $d$-space, measures centrality by a number between 0 and 1, while satisfying certain postulates regarding invariance, monotonicity, convexity and continuity. Accordingly, numerous notions of multivariate depth have been proposed in the literature, some of which are also robust against extremely outlying data.
The departure from classical Mahalanobis distance does not come without cost.
There is a trade-off between invariance, robustness and computational feasibility.
In the last few years, efficient exact algorithms as well as approximate ones have been constructed and made available in \texttt{R}-packages.
Consequently, in practical applications the choice of a depth statistic is no more restricted to one or two notions due to computational limits; rather often more notions are feasible, among which the researcher has to decide.
The article debates theoretical and practical aspects of this choice, including invariance and uniqueness, robustness and computational feasibility.  Complexity and speed of exact algorithms are compared.
The accuracy of approximate approaches like the {random Tukey depth} is discussed as well as the application to large and high-dimensional data.
{Extensions to local and functional depths and connections to regression depth are shortly addressed.}
\end{abstract}

\indent\\


\newpage

\section{Introduction}\label{sec1}
Many statistical tasks involve measures of centrality, identifying the center of a data cloud and measuring how close a given data point is to the center. In a probabilistic setting, one is interested in the question how central a point lies in a probability distribution. The opposite of centrality is outlyingness. Classical multivariate statistics measures outlyingness by the Mahalanobis distance. This is the usual Euclidean distance applied to `sphered' or `whitened' data, being transformed by a center point and a scatter matrix. Since the early 1990's more general statistics have been developed for measuring centrality and outlyingness of data in $\IR^d$ as well as for identifying central regions of a data cloud, consisting of points having at least a given centrality.
Though depth originates from data analysis and has been principally introduced for empirical distributions, most notions of depth allow for a \emph{population version}, that is, can be defined for general probability distributions beyond empirical ones.

In general, a ($d$-variate) \emph{depth function} is a function $D:(\bmy, P) \mapsto [0,1]$, for $\bmy\in \IR^d$ and $P$ from some class $\cal P$ of $d$-variate probability distributions, that satisfies several postulates regarding invariance, monotonicity, convexity and continuity.
$D(\bmy|\bmX)$ will be written in place of $D(\bmy|P)$, where $\bmX$ denotes a random variable distributed as $P$.
An often-quoted set of such postulates has been given in \cite{ZuoS00a}. Here a slightly terser one is used, which is due to \cite{Dyckerhoff02}: $D$ is a depth function if it is invariant against $\IR^d$-transformations in some class $\cal T$, null at infinity, monotone decreasing on rays from its maximum, and upper continuous. Formally, for $\bmy\in \IR^d$ and $P\in \cal P$,
\begin{itemize}
  \item \textbf{$\cal T$-Invariance:} $D(T(\bmy)|T(\bmX))=D(\bmy|\bmX)$ for all $T\in{\cal T}$,
  \item \textbf{Null at infinity:} $\lim_{\left\|\bmy\right\|\rightarrow\infty}D(\bmy|\bmX)=0$\,.
  \item \textbf{Monotone on rays:} If a point $\bmy^*$ has maximal depth, 
  that is \\ $D(\bmy^*|\bmX)=\max_{\bmy\in \IR^d}D(\bmy|\bmX)$\,
  then for any $\bmr$ in the unit sphere of $\IR^d$ the function $\gamma\mapsto D(\bmy^*+\gamma \bmr|\bmX)$ does not increase with $\gamma>0$\,.
  \item \textbf{Upper semicontinuous:} The upper level sets  $D^\alpha=\{\bmz\in \IR^d | D(\bmz|\bmX) \geq \alpha\}$ are closed for all $\alpha\in[0,1]$\,.
\end{itemize}
Any point $\bmy^*$ that has maximum depth is called a \textit{median}.
The postulates imply that the level sets (= \emph{central regions}) $D^\alpha$, $\alpha\in]0,1]$,
{are bounded. Monotonicity on rays means that they are starshaped about $\bmy^*$, hence nested.}
Moreover, if $\bmX$ is centrally symmetric about some $\bmz^*\in \IR^d$, then any depth function
yields $z^*$ as a median.
Recall that $\bmX$ is \emph{centrally symmetric} about $z^*$ if $\bmX-\bmz^*$  has the same distribution as $\bmz^*-\bmX$.
If the level sets are convex, $D$ is a \emph{quasi-concave depth function}.
Mostly, ${\cal T}$ is specified as the class of affine transformations of $\IR^d$, but other classes of transformations are possible and of practical interest.

Central regions are sometimes parameterized by their probability content,
\begin{equation} \label{probcentral}
D_{\rm{\beta}}(\bmX) =\bigcap_{\alpha \in A(\beta)} D^{\alpha}(\bmX) \,,\quad \text{where} \quad A(\beta)= \{\alpha : P[D^{\alpha}(\bmX)] \ge \beta\}\,.
\end{equation}

If $P$ is the empirical distribution on a set $\{\bmx_1, \dots, \bmx_n\}$ of data points, the depth function is mentioned as a \emph{multivariate data depth} and written $D(\bmy|\bmx_1, \dots, \bmx_n)$.

Well-known examples of depth functions are the \emph{halfspace depth} \citep{Tukey75}, which is also called  \emph{Tukey}  or  \emph{location depth},  the \emph{zonoid depth} \citep{KoshevoyM97b}, the \emph{spatial} \citep{Serfling02}, \emph{projection} \citep{Liu92,ZuoS00a}, \emph{simplicial} \citep{Liu90} and  \emph{simplicial volume} \citep{Oja83} \emph{depths}. A more recent notion is the \emph{$\beta$-skeleton depth} \citep{YangM17}, which includes the \emph{lens depth} \citep{LiuM11} and the \emph{spherical depth} \citep{ElmoreHX06} as special cases; see Section~\ref{sec2} below.
These depth functions differ in their analytical properties and computational feasibility.
When it comes to applications the problem arises which of these depth notions should be employed in a given situation.

Depth statistics have been used in numerous and diverse tasks of which we can mention a few only. \cite{LiuPS99} provide an introduction to some of them.
Given data in $\IR^d$, central regions are \emph{set-valued statistics} {that provide multivariate ranks and quantiles}. They are used
to {describe} and compare multivariate distributions w.r.t.
 location, dispersion, and shape, {e.g. by a bagplot \citep{RousseeuwRT99},}
to identify {outliers} of a distribution \citep{ChenDPB09},
to {classify} and {cluster} data \citep{Hoberg00, LangeMM14a},
to {test} for multivariate scale and symmetry \citep{Dyckerhoff02, DyckerhoffLP15}. Also,
to {measure} multidimensional risk \citep{CascosM07},
to handle constraints in {stochastic optimization} \citep{MoslerB14},
{and for quantile regression \citep{Chakraborty03,HallinPS10}},
among others.

Actually, a plethora of depth notions can be defined that satisfy the postulates, and many different notions have already been proposed in the literature.
Classical multivariate statistics based on {Mahalanobis distance} always yields elliptical central regions, which correspond to a model assumption of
elliptically symmetric probability distributions. In contrast, other depth statistics adapt better to non-symmetric distributions. Some depth statistics are also robust against possibly contaminated data. The departure from classical Mahalanobis distance does not come without cost. The computational load cannot be neglected, gaining weight with the number $n$ of data and, even more, with dimension $d$. Therefore, until recently, due to computational infeasibility the use of most depth statistics was limited to small $n$ and $d$ in applications, and the choice of a proper depth statistic  in practice was restricted to very few notions.

But in the last few years, efficient algorithms have been constructed. Procedures for many notions of multivariate and functional data depth have been made available in \texttt{R}-packages and applied to various tasks; see \cite{PokotyloMD19} for \texttt{ddalpha}, \cite{depth} for \texttt{depth}, \cite{HubertRS15} for \texttt{mrf.Depth},
\cite{KosiorowskiZ14} for \texttt{DepthProc},
and \cite{FebreroBandeOF12} for \texttt{fda.usc}. For details see Section~\ref{sec5.3} below.

These packages allow, together with the secular increase in computing power, the numerical treatment of depth statistics in applications having realistic sizes of $n$ and $d$. Consequently, the choice of a depth statistic is not any more restricted to one or two notions due to computational limits, but rather often to more notions of depth among which {the statistician has} to decide.

Also, depth notions for functional data have raised considerable interest in the recent literature. Most of them build on depths for finite dimensional data, so that their properties (including computability) depend on that of the multivariate depths involved.

This article discusses aspects and general principles that guide in this choice. They are useful in the construction of depth-based statistical procedures as well as in practical applications when several notions of depth appear to be computationally feasible.
{Depth statistics are primarily viewed as a data-analytic tool, without assuming a specific data generating process; a view that is directly associated with Tukey's original idea.
In this locational sense depth has been used in a huge amount of statistical methodology and applications so far.
The objective is neither giving a general theory of depth statistics nor a historical account of its development. We rather aim at a comparative evaluation of multivariate depth notions that are broadly used in many statistical procedures regarding their analytical properties and practical computability. For comparison nine widely used notions of depth have been selected (each having hundreds of references in the literature), plus two more special ones, which are constructed for special types of data. According to our data-analytic approach peeling depth is included despite its missing population version.
The focus is in global depths in $d$-space; extensions to local and functional depths and connections to regression and tangent depth are shortly covered.}

In Section~\ref{sec2}  eleven {selected} depth statistics are reviewed. Section~\ref{sec3} discusses their analytical properties: invariance, depth-trimmed regions, uniqueness of the underlying distribution, and continuity. Section~\ref{sec4} is about possible specifics of the data, such as symmetry of the generating law, the existence of outliers and the missing of data.
Section~\ref{sec5} treats the computational feasibility of the depth notions; times for their exact computation are compared, approximate procedures discussed, and the approximation error of the random Tukey depth \citep{CuestaAN08b} is calculated. {Existing implementations of exact and approximate procedures are shortly surveyed.} Then, dealing with large and high-dimensioned data is discussed.
Section~\ref{sec6} addresses extensions to local and functional depths {as well as connections to regression and tangent depth}.
Section~\ref{sec7} concludes with practical guidelines.

\section{Some depth statistics}\label{sec2}
Many depth notions have been proposed in the literature. Not all of them satisfy the above postulates. Here, {nine most popular ones are reviewed and, additionally, two more recent notions that apply also to ordinally scaled distances.}
For some of these depth notions, $\bmX$ has to satisfy moment conditions or other obvious restrictions.

\begin{itemize}
\item \textbf{Mahalanobis depth} \citep{Mahalanobis36}:
\begin{equation}\label{gMahdepth}
D_{\rm {Mah}}(\bmy|\bmX) =
         \left( 1 + ||\bmy-\bmm_\bmX||^2_{\boldsymbol{\Sigma}_\bmX}\right)^{-1}
\end{equation}
is called \emph{(moment) Mahalanobis depth}. Here, $\bmm_\bmX$ and $\Sigma_\bmX$  denote the expectation vector
and the covariance matrix of $\bmX$,  and  $||\bmz||_{\boldsymbol{\Sigma}_\bmX}^2= \bmz^T \boldsymbol{\Sigma}^{-1}_\bmX \bmz$ is the \emph{Mahalanobis norm} of $\bmz \in \IR^d$.

    \item \textbf{$\IL_p$ depth} \citep{ZuoS00a}:
\begin{equation}\label{L2depth}
D_{\rm {\IL_p}}(\bmy|\bmX) = \left( 1 + E [||\bmy-\bmX||_p]\right)^{-1} \,,
\end{equation}
$1\le p<\infty$. The depth \emph{$D_{\rm {\IL_p}}$} is based on the expected
outlyingness of a point, as measured by the $\IL_p$ distance.
For $p=2$ it is also mentioned as \emph{Euclidean depth}.

  \item \textbf{Halfspace depth ( = location depth = Tukey depth)} \citep{Tukey75, DonohoG92}:
  \begin{equation}\label{Hdepth}
D_{\rm H}(\bmy|\bmX) = \inf \{P[\bmX\in H]  :  H \; \mbox{closed halfspace,}  \quad \bmy \in H\}\,.
\end{equation}

    \item \textbf{Projection depth} \citep{Liu92,ZuoS00a}:
\begin{equation}\label{projdepth}
D_{\rm {Proj}}(\bmy|\bmX) =
         \left(1 + \sup_{\bmp\in S^{d-1}}
         \frac {|\langle \bmp,\bmy \rangle
                - \med(\langle \bmp,\bmX \rangle)|}
               {\mad(\langle \bmp,\bmX \rangle)}
         \right)^{-1} \, ,
\end{equation}
where $\med(V)$ denotes the median of a univariate random variable $V$,
and $\mad(V)=\med(|V-\med(V)|)$ its median absolute
deviation from the median.

\item \textbf{Simplicial depth} \citep{Liu90}:
    \begin{equation}\label{Simdepth}
D_{\rm {Sim}}(\bmy|\bmX) = P\left[\bmy \in \,\conv(\{ \bmX_{1},\ldots ,\bmX_{{d+1}} \})\right] ,
\end{equation}
where $\bmX_{1},\ldots ,\bmX_{{d+1}}$ are i.i.d.\ copies of $\bmX$, and $\conv$  means convex hull.

    \item \textbf{Simplicial volume depth ( = Oja depth)} \citep{Oja83,ZuoS00a}:
    For any points $\bmv_1, \dots, \bmv_{d+1} \in\IR^d$, the convex hull
$\conv(\{\bmv_{1},\ldots ,\bmv_{{d+1}}\})$ has $d$-dimensional volume
$$\vol_d(\conv \{ \bmv_1,\ldots,\bmv_{d+1}\})=
\frac 1{d\,!}\ |\mbox{det}((1,\bmv_{1}^\top)^\top,\ldots,(1,\bmv_{d+1}^\top)^\top)|\,.$$
The {\em simplicial volume depth} or \emph{Oja depth} is defined by
\begin{equation}\label{OjaDepth}
D_{\rm {Oja}}(\bmy|\bmX) =\left( 1 + {E \left[\vol_d(\conv\{\bmy,\bmX_{1},\ldots ,\bmX_{d}\})\right]} \right)^{-1}  ,
\end{equation}
where $\bmX_1, \dots, \bmX_d$
are independent copies of $\bmX$.

    \item \textbf{Zonoid depth} \citep{KoshevoyM97b}:
  For $0<\alpha\le 1$,
\begin{align}\label{inttr1}
D_{\rm {Zon}}^{\alpha}(\bmX)= \Bigl\{
& E[\bmX\, g({ \bmX})]\; :\; g:\IR^d \rightarrow \left[0,1/\alpha \right]
 \enspace \\ & \mbox{measurable and}\,\,E[g( \bmX)]=1 \Bigr\} \nonumber
\end{align}
is the {\em zonoid} $\alpha$-{\em region} of $\bmX$. For $\alpha = 0$
set $D_{\rm {Zon}}^0(\bmX) = \IR^d$. The \emph{zonoid depth} is defined as
\begin{equation}\label{ZonoidDepth}
   D_{\rm {Zon}}(\bmy|\bmX) = \sup \{\alpha : \bmy\in D_{\rm {Zon}}^{\alpha}(\bmX)\}
\end{equation}
For an empirically distributed $\bmX$ the zonoid depth then is calculated as
\begin{align}\label{ZonoidDataDepth}
D_{\rm {Zon}}(\bmy|\bmX) =\sup \Bigl\{ & \alpha  \, : \,
\alpha  \lambda_i \le  1/n,
{ \bmy}=\sum_{i=1}^n\lambda_i{ \bmx}_i,  \,\, \\
& \sum_{i=1}^n\lambda_i=1, \,\,
\lambda_i\ge 0 \;\, \forall i \Bigr\}\,. \nonumber
\end{align}
    \item \textbf{Spatial depth} \citep{Serfling02}:
    The \emph{spatial depth} is defined as
\begin{equation}\label{SpatialDepth}
   D_{\rm {Spa}}(\bmy|\bmX) = 1- \left|\left|E\left[\frac{\bmy-\bmX}{||\bmy-\bmX||}\right]\right|\right|\,,
   \end{equation}
where $0/0=0$.
    \item \textbf{Lens depth} \citep{LiuM11}:
A function $v:\IR^d\times \IR^d\to \IR_+$ is a \emph{dissimilarity function} if it is symmetric, $v(\bmx, \bmy)= v(\bmy, \bmx)$, and vanishes if and only if $\bmx$ and $\bmy$ are the same, $v(\bmx, \bmy)= 0 \Leftrightarrow \bmx= \bmy$. Note that every distance is a dissimilarity function.
Assume that for any three values $\bmx,\bmy,\bmz$ of a random vector $\bmX$ in $\IR^d$ the following information is given: One of these points, say $\bmx$, is more similar (= less dissimilar) to each of the two other points than these are among themselves, i.e.
\begin{equation}\label{ordinalInfo}
v(\bmx, \bmy)<v(\bmy, \bmz) \quad \text{and} \quad v(\bmx, \bmz)<v(\bmy, \bmz),
\end{equation}
in symbols $\bmx\lhd(\bmy,\bmz)$. It means that $\bmx$ is the ``most central'' among the three points.
This information will be mentioned as \emph{ordinal dissimilarity information} on $\bmX$.

Given such ordinal dissimilarity information, the \emph{lens depth} can be defined as follows \citep{KleindessnerVL17}:
\begin{equation}\label{LensDepth}
  D_{\rm {Lens}}(\bmy|\bmX) = P[\bmy\lhd(\bmX_1,\bmX_2)]\,,
\end{equation}
where $\bmX_1$ and $\bmX_2$ are independent copies of $\bmX$. If dissimilarity is measured by the Euclidean distance one gets
\begin{equation}\label{LensDepthEucl}
  D_{\rm {Lens}}^{\rm{Eucl}}(\bmy|\bmX) = P[\max \{ ||\bmy-\bmX_1||, ||\bmy-\bmX_2|| \} < ||\bmX_1-\bmX_2||] \,.
\end{equation}
\item $\beta$\textbf{-skeleton depths} \citep{YangM17}:
Depending on a parameter $\beta\ge 1$ \cite{YangM17} introduced the
\emph{$\beta$-skeleton depths},
\begin{align}\label{SkeletonDepth}
	  D_{\rm {Skel\beta}}(\bmy|\bmX) = P[&||\bmy - \bmX_{1\beta 2}|| < \beta/ 2 ||\bmX_1-\bmX_2|| \\
  \text{and} \, & ||\bmy - \bmX_{2\beta 1}|| <  \beta/2||\bmX_1-\bmX_2||]\,, \nonumber
\end{align}
where $\bmX_{1\beta 2} = \beta/2 \bmX_1 + (1-\beta/2) \bmX_2 $. With $\beta=2$ the \emph{Euclidean lens depth} (\ref{LensDepthEucl}) is obtained, and with $\beta=1$ the so called \emph{spherical depth} \citep{ElmoreHX06},
\begin{equation}\label{SphericalDepth}
  D_{\rm {Sph}}(\bmy|\bmX) =  P\left[||\bmy -  (\bmX_{1} + \bmX_{2})/2|| <  ||\bmX_1-\bmX_2|| /2 \right]\,.
  \end{equation}
    \item \textbf{Convex hull peeling depth ( = onion depth)} \citep{Barnett76, Eddy81}: The convex hull peeling depth is defined only for an empirical distribution of $\bmX$, say on the finite set $S_\bmX$.
        Consider the sequence of closed convex sets (= \emph{onion layers})
\begin{equation}\label{ConvexPeeling}
  C_1(\bmX)=\conv(S_\bmX)\,, \quad C_{j+1}(\bmX)= \conv(S_\bmX\cap \interior\, C_j(\bmX))\,, \quad j=1,2,\dots\,.
\end{equation}
Here  $\interior$  denotes the interior of a set. Define the \emph{convex hull peeling} (or \emph{onion}) \emph{depth} of $\bmy$ as the smallest index $j$ at which $\bmy\in C_j$, i.e.
\begin{equation}\label{ConvexHullDepth}
  D_{\rm{Onion}}(\bmy|\bmX)= \sum_{j\ge1} \bm1_{\interior\, C_j(\bmX)}(\bmy)\,,
\end{equation}
with $\bm1_S$ denoting the indicator function of a set $S$.
    \end{itemize}

\begin{figure}[!t]
\begin{center}
\begin{tabular}{ccc}
\quad {\small Mahalanobis depth} & \quad {\small $\IL_2$ depth} & \quad {\small Halfspace depth} \\
\includegraphics[width=0.31\textwidth,trim = 0cm 0.5cm 1cm 1.5cm,clip=true,page=2]{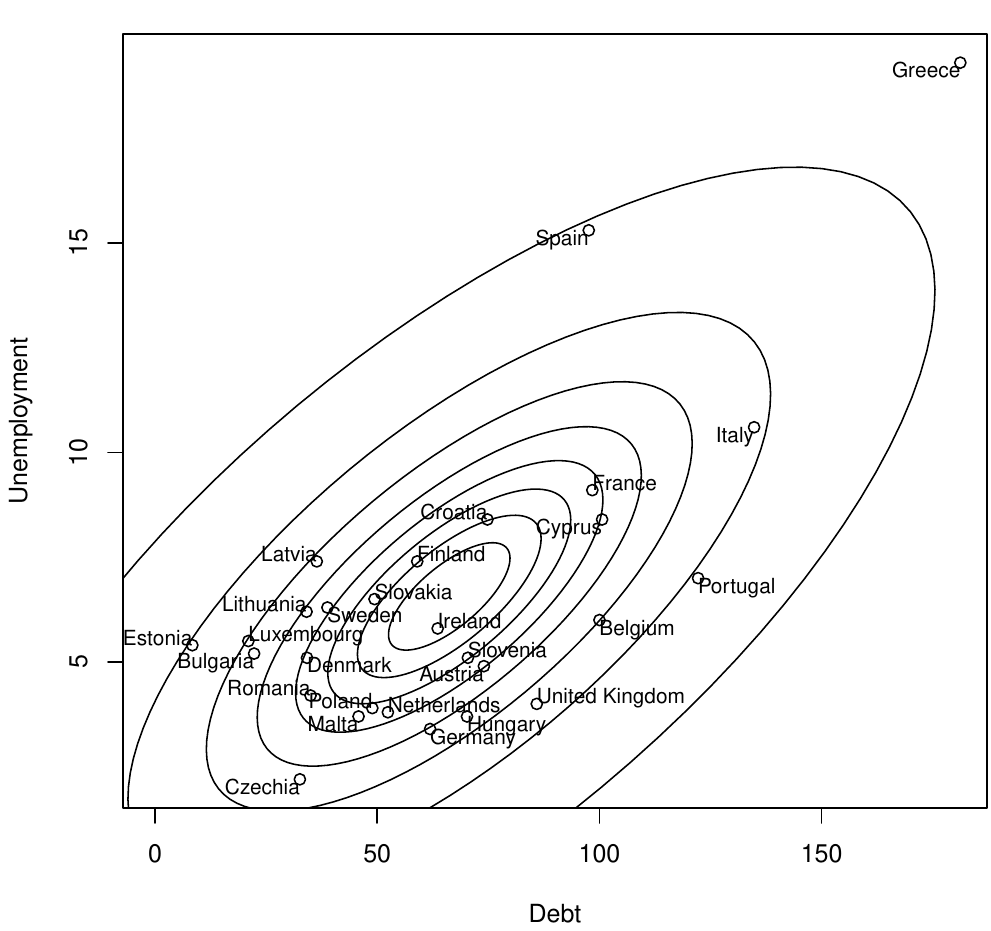} & \includegraphics[width=0.31\textwidth,trim = 0cm 0.5cm 1cm 1.5cm,clip=true,page=12]{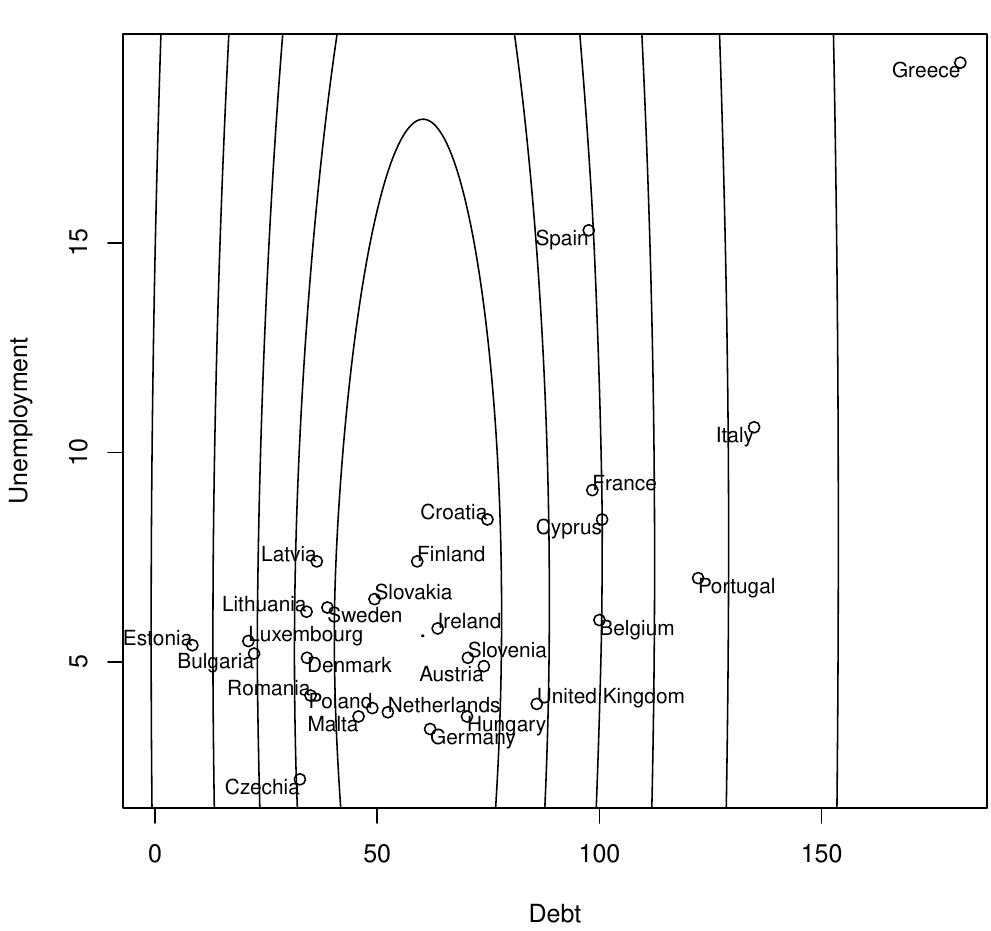} & \includegraphics[width=0.31\textwidth,trim = 0cm 0.5cm 1cm 1.5cm,clip=true,page=9]{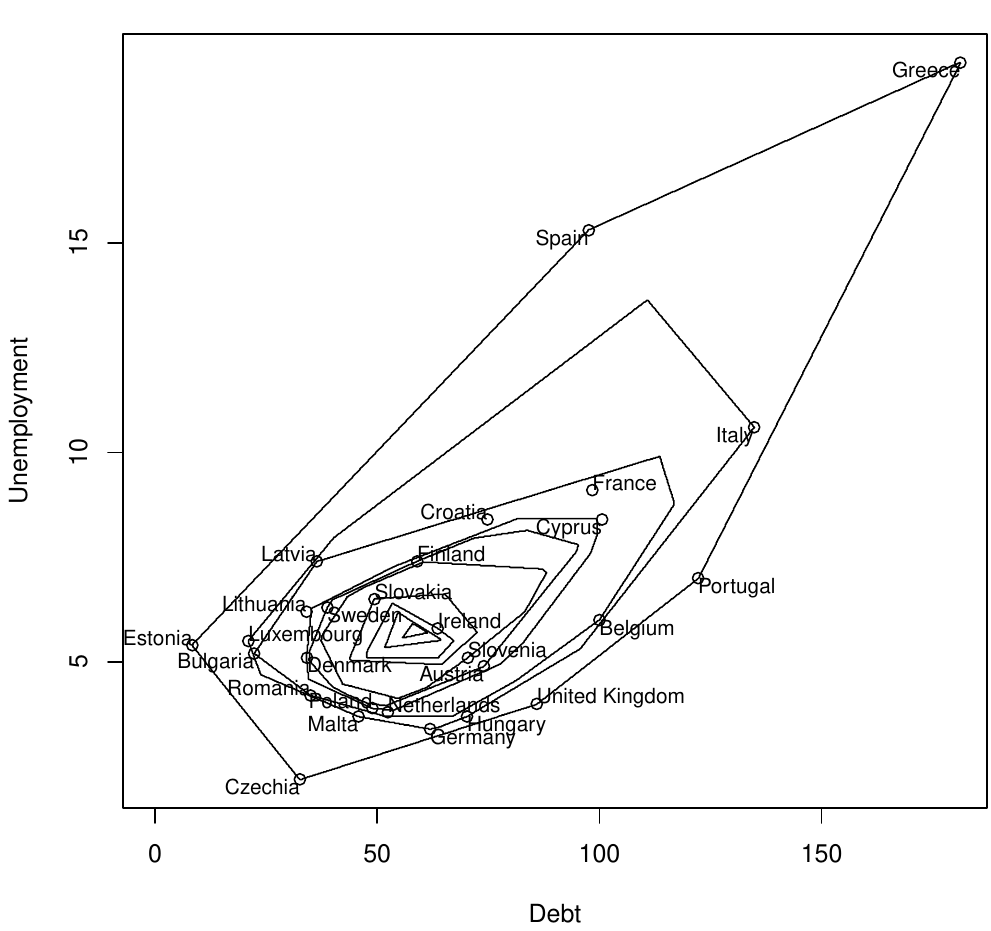} \\
\quad {\small Projection depth} & \quad {\small Simplicial depth} & \quad\,\,{\small Simplicial volume depth} \\
\includegraphics[width=0.31\textwidth,trim = 0cm 0.5cm 1cm 1.5cm,clip=true,page=5]{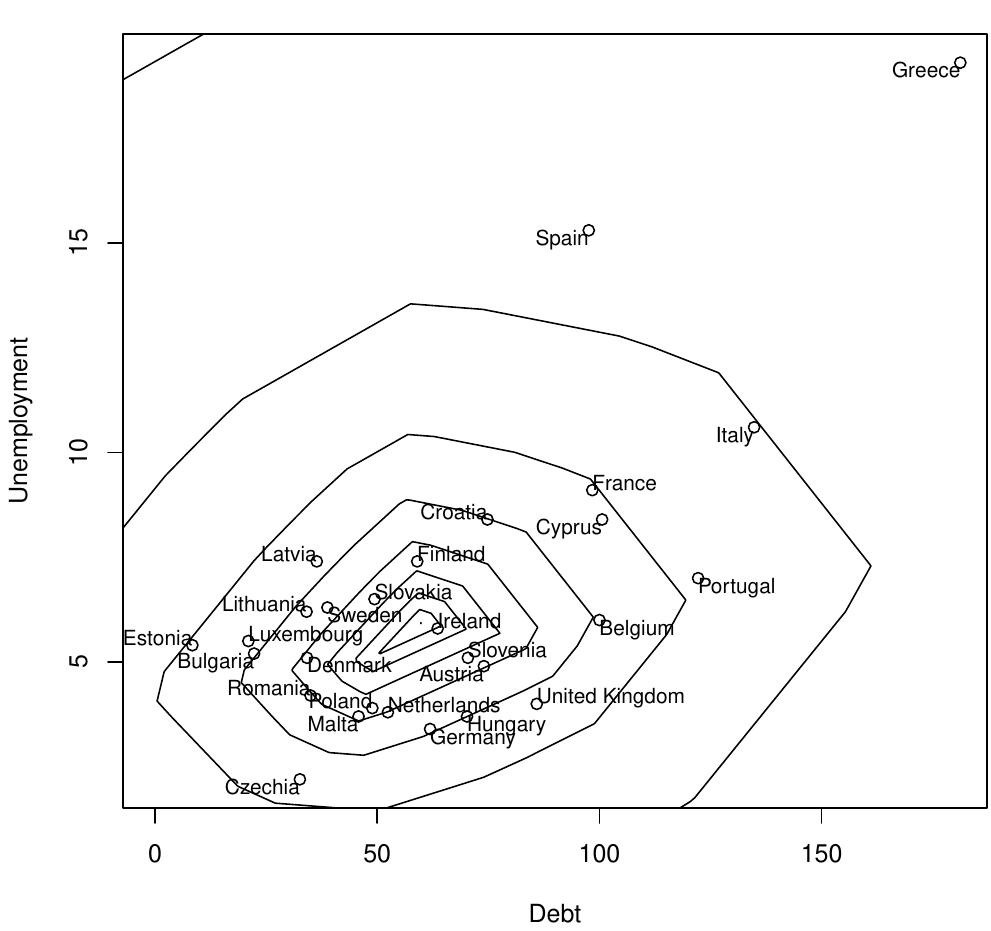} & \includegraphics[width=0.31\textwidth,trim = 0cm 0.5cm 1cm 1.5cm,clip=true,page=7]{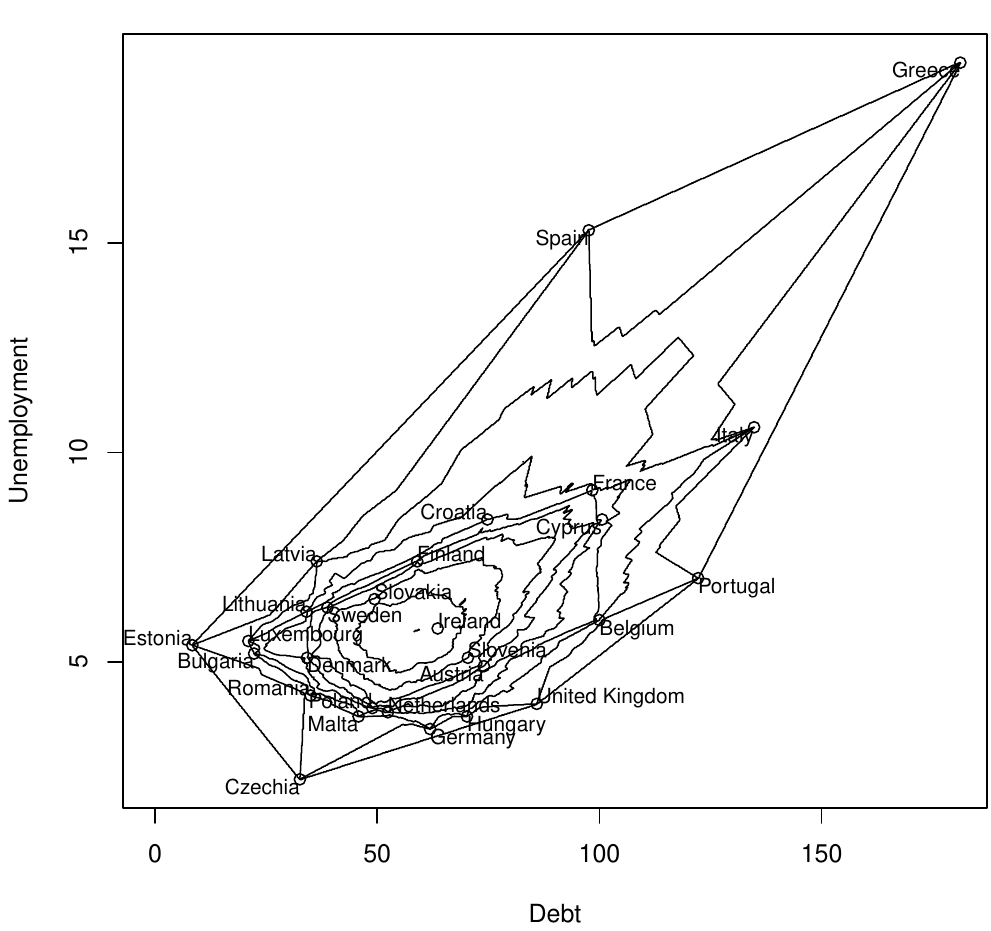} & \includegraphics[width=0.31\textwidth,trim = 0cm 0.5cm 1cm 1.5cm,clip=true,page=8]{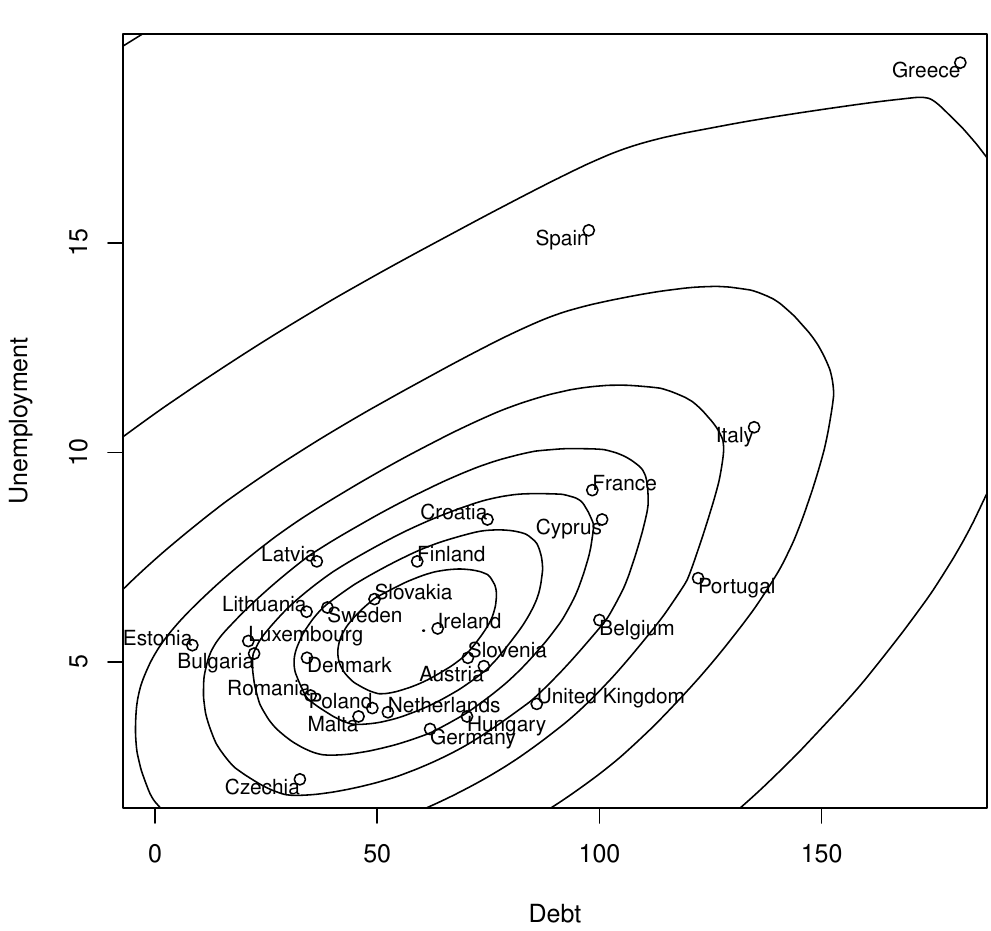} \\
\quad {\small Zonoid depth} & \quad {\small Spatial depth} & \quad {\small Onion depth} \\
\includegraphics[width=0.31\textwidth,trim = 0cm 0.5cm 1cm 1.5cm,clip=true,page=4]{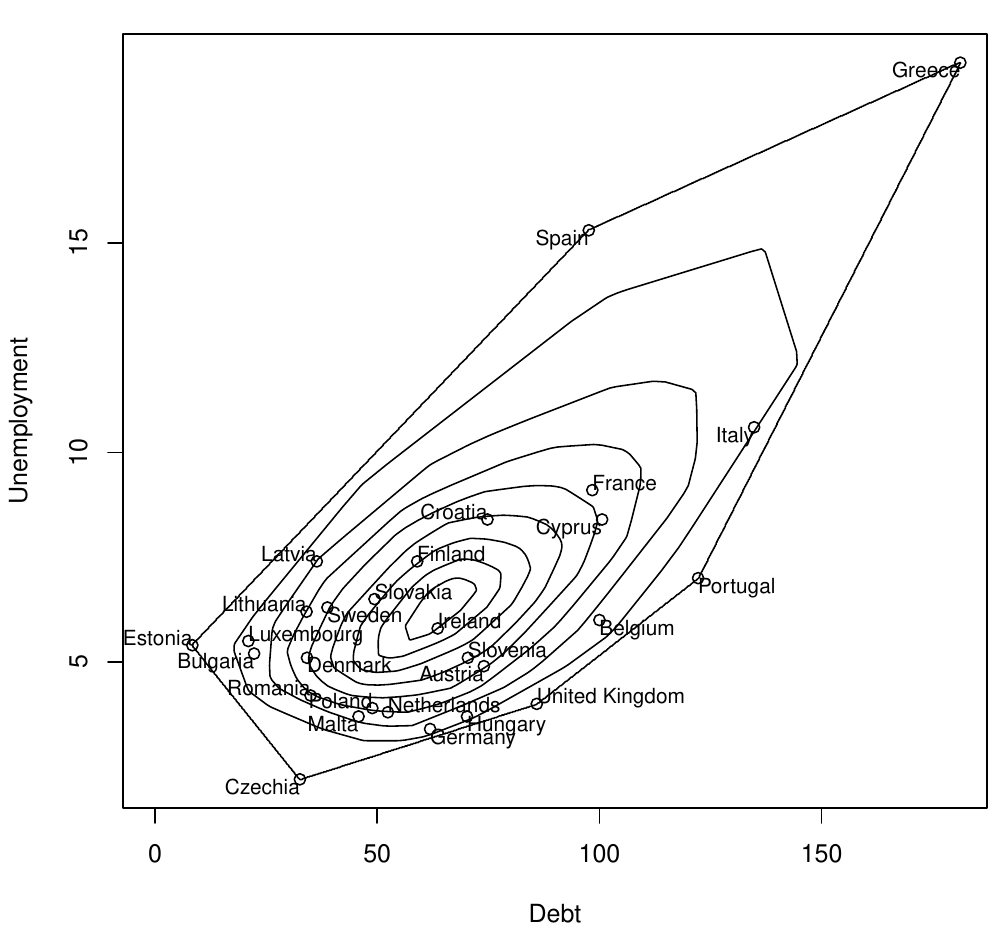} & \includegraphics[width=0.31\textwidth,trim = 0cm 0.5cm 1cm 1.5cm,clip=true,page=6]{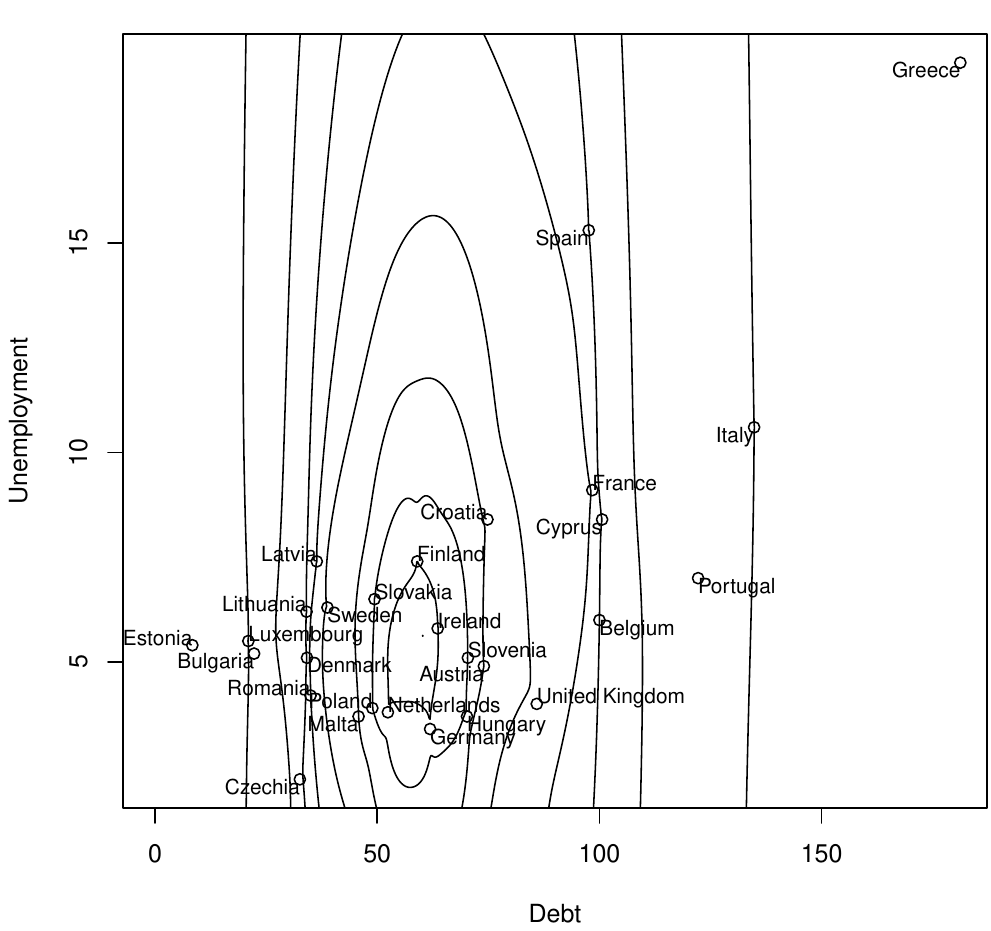} & \includegraphics[width=0.31\textwidth,trim = 0cm 0.5cm 1cm 1.5cm,clip=true,page=10]{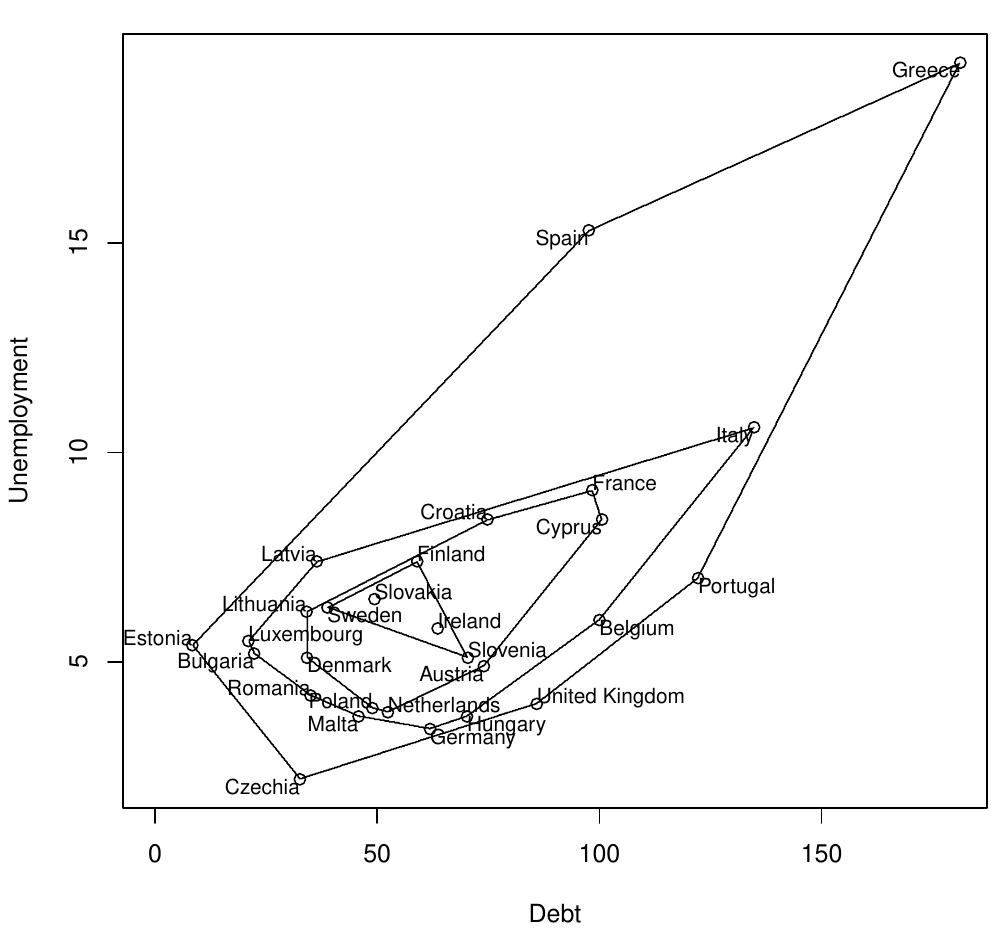} \\
\end{tabular}
\end{center}
\caption{Central regions of EU data on unemployment and public debt in 2018 (Source: EUROSTAT), for nine notions of depth.}\label{figure:EU1}
\end{figure}

Figure~\ref{figure:EU1}
exhibits, for nine notions of depth, central regions of
bivariate macroeconomic data (unemployment and public debt in 2018) of all 28 countries of the European Union. For lens depth see Figure~\ref{figure:EU2}.


\section{Relevant properties}\label{sec3}

As is seen from Figure~\ref{figure:EU1},
different depths yield different central regions.
Consequently, when these depths are employed, e.g.\ to find outliers or to classify data, different results will arise.
Therefore one has to distinguish the specific aspects of the various depths. In the sequel properties of the depth notions are discussed {for dimension $d\ge 2$}. (Note that for reasons of practicality and comparison these properties are not always given in their most general form.)
Many properties of Mahalanobis, {$\IL_p$}, halfspace, simplicial, projection, and Oja depth are demonstrated in \cite{ZuoS00a}. Particularly,
see \cite{DonohoG92} for halfspace depth, \cite{ZuoS00a} for $\IL_p$ depth, \cite{Serfling02} for spatial depth,
\cite{Mosler02a} for zonoid depth, \cite{Zuo03} for projection depth. For $\beta$-skeleton and lens depth refer to
\cite{LiuM11} and \cite{YangM17}, and for onion depth to
\cite{DonohoG92} and \cite{LiuPS99}.

\subsection{Invariance of depth statistic}\label{sec3.1}
A principal aspect of any statistical procedure is invariance: Which features of the data shall the statistic ignore, and which shall it reflect?

Table~\ref{tab:invariance} specifies relevant classes $\cal T$ of transformations of $\IR^d$ to which a depth function $D$ may be invariant, $D(T(\bmy)|T(\bmX))=D(\bmy|\bmX)$ for all $T\in {\cal T}$.

\begin{table}[h!]
\caption{Classes of transformations for invariance of depth functions.}\label{tab:invariance}
\begin{center}
  \begin{tabular}{|l|l|l|l|}
  \hline
  $\cal T$ & Name & Transformation & \\
  \hline
  $\cal T_{A}$ & affine& $\bmx\mapsto \bmA \bmx +\bmb$ & $\bmA\in\IR^{d\times d}$ regular, $\bmb\in \IR^{d}$ \\
  $\cal T_{O}$ & orthogonal & $\bmx\mapsto \bmA \bmx$ & $\bmA\in\IR^{d\times d}, \bmA \bmA^T = I_d$ \\
  ${\cal T}_{T}$ & translation & $\bmx\mapsto \bmx +\bmb$ &  $\bmb\in \IR^{d}$ \\
  ${\cal T}_{cSc}$ & coordinate-wise scaling & $\bmx\mapsto \boldsymbol{\Lambda} \bmx$ & $\boldsymbol{\Lambda}=\it{diag}(\lambda_1,\dots, \lambda_d), \lambda_j>0$\\
  ${\cal T}_{uSc}$ & uniform scaling & $\bmx\mapsto \lambda\bmx$ & $\lambda> 0$\\
${\cal T}_{Com}$ & combinatorial &  $T$ combinatorial transf. & \\
${\cal T}_{OD}$ & ordinal dissimilarity & $T$ stretching transf. & \\
  \hline
  \end{tabular}
\end{center}
\end{table}

A depth is \emph{combinatorially invariant} if it is invariant against {combinatorially equivalent} transformations of its arguments $\bmy, \bmx_1, \dots, \bmx_n$.
A transformation of data is mentioned as \emph{combinatorially equivalent} if, besides renumbering,
none of the data crosses a hyperplane spanned by $d$ other data points;
more precisely, if the set of minimal Radon partitions of the data remains unchanged; see Section~4.4. in \cite{Mosler02a}.
{(A \emph{Radon partition} of a data matrix $[\bmx_1, \dots, \bmx_n]^T$ divides the set of row indices,
$K=\{1,\dots,n\}$, into three disjoint subsets $K_+,K_-$, and $K_0$ so that
the convex hulls of the pertaining points intersect, i.e.\ $\conv \{\bmx_j : j\in K_+\} \cap  \conv \{\bmx_j : j\in K_-\} \not= \emptyset.$
It is called \emph{minimal} if
no other Radon partition of $[\bmx_1, \dots, \bmx_n]^T$ has smaller support $K_+ \cup K_-$.)}
Then, in particular, any point that lies on the convex hull border of the data can be moved far away from the data cloud without changing its depth. Consequently, a combinatorially invariant depth is very robust against outlying data, while it is \emph{not} useful in {identifying outliers}.
Note that {combinatorial invariance} is defined on empirical distributions only. A depth that is combinatorially invariant is named a \emph{combinatorial depth}.

Halfspace depth, simplicial depth, and onion depth are combinatorially invariant; see e.g. Cor.~4.12 in \cite{Mosler02a}.
In contrast, other depths are not combinatorially invariant, as they use distances in $\IR^d$:   Mahalanobis, $\IL_2$, lens,
 projection, and simplicial volume depth. Also the zonoid depth as well as the weighted-mean depths \citep{DyckerhoffM11}
 use the metrical structure of $\IR^d$.

A \textit{stretching transformation} is a transformation of the data that leaves the ordinal dissimilarity information unchanged, {in particular by an increasing transform of the dissimilarity function}.
The lens depth with general dissimilarity function is \textit{ordinal dissimilarity invariant}.
Note that, if the data is metrically scaled and $v$ is the Euclidean (resp.\ Mahalanobis) distance, then the lens depth is orthogonal and translation (resp.\ affine) invariant.

Many multivariate depths are \emph{affine invariant}, which is often considered as a standard requirement.
These depths are independent of any specific coordinate system in $\IR^d$. Examples are the Mahalanobis, halfspace, projection, simplicial, zonoid, and onion depths.
Affine invariance may hold only up to a positive scalar factor; this is mentioned as \emph{weak affine invariance}.

\textbf{Sphering the data: }Some depths are basically invariant only to orthogonal transformations, translations and uniform scaling, among them the
Euclidean, spatial, simplicial volume, and $\beta$-skeleton depths.
These depths can be made affine invariant by a scatter matrix transform; {see e.g.\ \cite{Serfling10}}.

A \emph{scatter matrix} $\bmR_{\bmX}$ (also called \emph{scatter functional}) is a symmetric positive definite
$d\times d$ matrix that depends continuously (in weak convergence) on
the distribution of $\bmX$ and measures its spread in an affine equivariant way.
The latter means that
\begin{equation}\label{affequdisp}
\bmR_{\bmA\bmX  + \bmb}= \lambda_{\bmX,\bmA,\bmb} {\bmA}\bmR_{\bmX}{\bmA}^T \quad
\mbox{holds for any}\; \bmA \; \mbox{of full rank and any} \; \bmb ,
\end{equation}
with some $\lambda_{\bmX,\bmA,\bmb}>0$.
The data is transformed as
\begin{equation}\label{sphering}
  \bmx \mapsto \bmR_\bmX^{-1/2}(\bmx - \theta(\bmX))\,,
\end{equation}
where $\bmR_\bmX$ is a scatter matrix and $\theta(\bmX)$ a location parameter.
This scatter matrix transformation is also mentioned as \emph{sphering} or \emph{whitening} the data.

\begin{figure}
\begin{center}
\begin{tabular}{ccc}
\quad {\small } & \quad {\small } & \quad {\small Affine invariant } \\
\quad {\small Lens depth} & \quad\,\,{\small Affine invariant lens depth} & \quad {\small simplicial volume depth} \\
      \includegraphics[width=0.31\textwidth,trim = 0cm 0.5cm 1cm 1.5cm,clip=true,page=11]{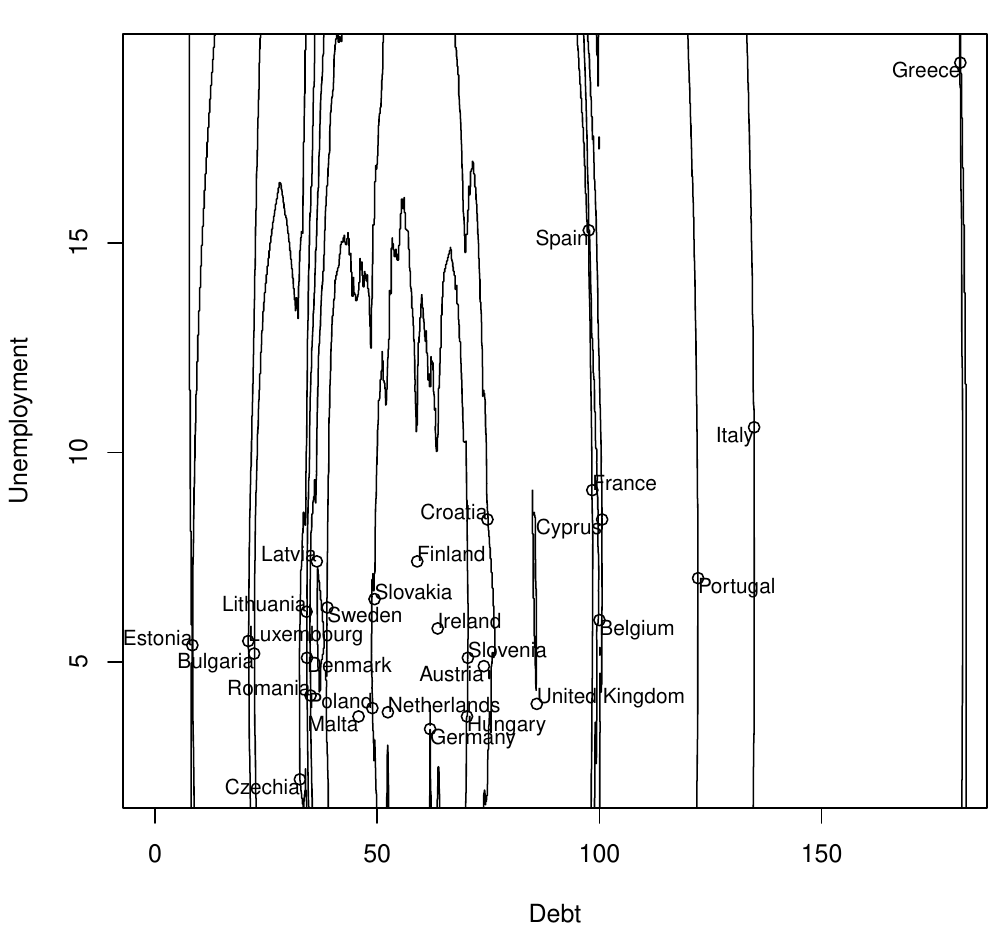} & \includegraphics[width=0.31\textwidth,trim = 0cm 0.5cm 1cm 1.5cm,clip=true,page=17]{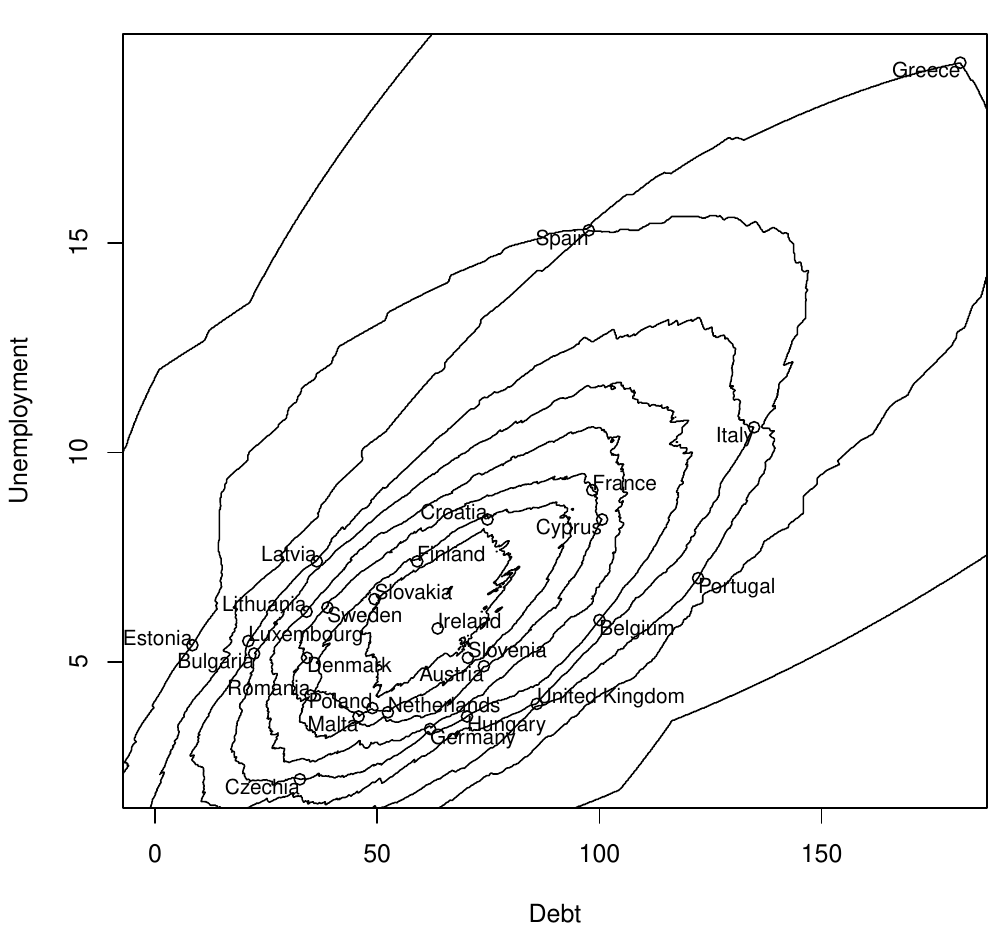} & \includegraphics[width=0.31\textwidth,trim = 0cm 0.5cm 1cm 1.5cm,clip=true,page=15]{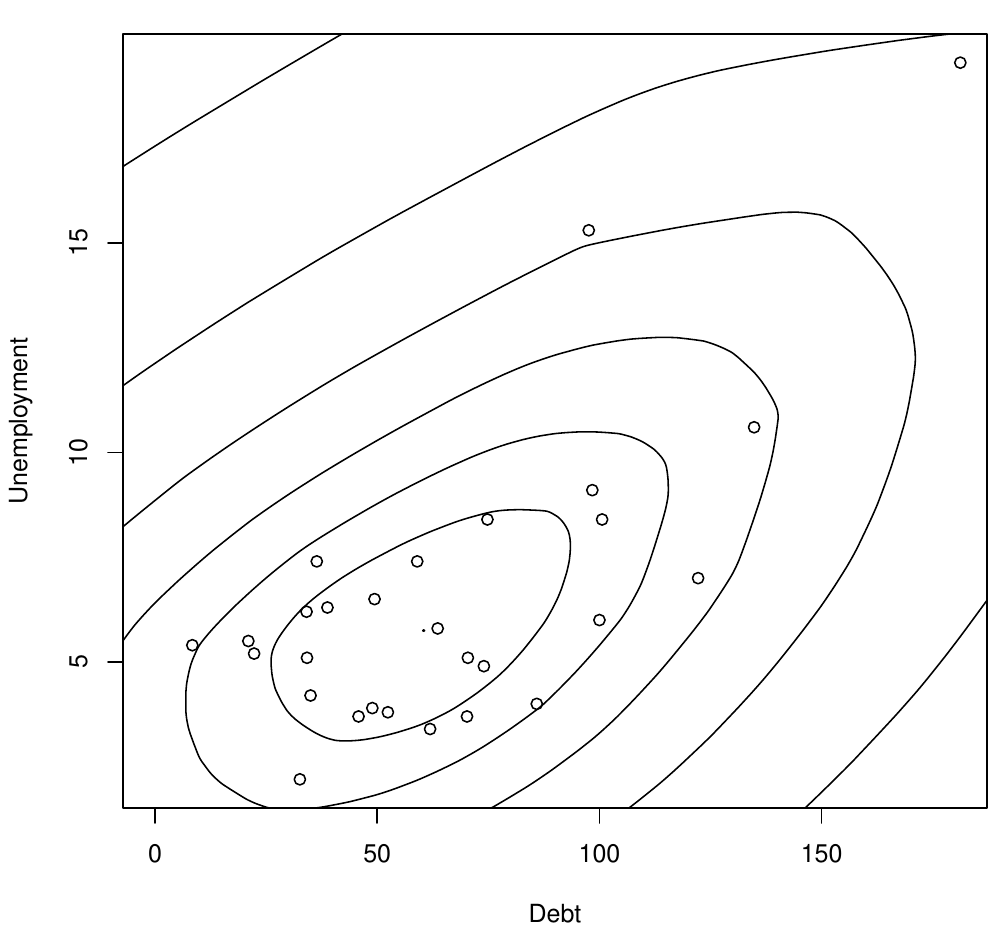}
\end{tabular}
\end{center}
\caption{Central regions of EU data on unemployment and public debt in 2018 (Source: EUROSTAT): lens depth, affine invariant lens depth, and affine invariant simplicial volume depth ($\bmR=$ covariance matrix).}\label{figure:EU2}
\end{figure}

E.g., the simplicial volume depth (\ref{OjaDepth}) is only orthogonal invariant. Its affine invariant version is given by
\begin{equation}\label{affineOja}
D^*_{\rm {Oja}}(\bmy|\bmX) =\left( 1 + \frac
{E \left[\vol_d(\conv\{\bmy,\bmX_{1},\ldots ,\bmX_{d}\})\right]}
{\sqrt{\mbox{det\,}\bmR_{\bmX}}}
\right)^{-1}  ,
\end{equation}
Similarly, the lens depth (\ref{LensDepth}) is made affine invariant. Central regions of lens depth, affine invariant lens depth, and affine invariant simplicial volume depth are shown in Figure~\ref{figure:EU2}.


Clearly, there are many ways to choose a matrix that satisfies (\ref{affequdisp}).
A most prominent example for $\bmR_{\bmX}$
is the covariance matrix ${\boldmath\Sigma}_{\bmX}$ of $\bmX$.
(Observe that the covariance matrix is positive definite, if  the convex hull of the support of $\bmX$ has full dimension.)
If robustness is an issue, proper choices for the scatter matrix $\bmR_{\bmX}$
are the {\it minimum volume ellipsoid}\/ (MVE) estimator, the
{\it minimum covariance determinant}\/ (MCD) estimator and similar
robust covariance estimators \citep{RousseeuwL87,LopuhaaR91}. A robust depth-based scatter estimate is proposed in \cite{PaindaveineVB2018}. See also Section~\ref{sec4.2} below.

\textbf{Different tasks:}
However, many classical multivariate procedures are less than affine invariant and, if combined with a depth statistic, do not ask for an affine invariant depth notion. The following task settings ask for different kinds of invariance.
\begin{itemize}
  \item Affine invariance: general independence of coordinate system, e.g. in estimating parameters, testing hypotheses, and outlier identification.
  \item Weak affine invariance (= affine invariance up to a scalar constant):
  independence of coordinate system up to a homogeneous scale change.
  \item Orthogonal 
  invariance: the set of Euclidean distances among pairs of points has to remain unchanged.
  \item  Translation invariance: pure location problems and problems that depend on the specific meaning of the coordinate axes (e.g. length and weight).
  \item Invariance to coordinate-wise  resp.\ uniform scaling: general dispersion problems resp. dispersion problems having a common measurement scale of coordinates (e.g. lengths of some object).
\item Ordinal dissimilarity invariance: problems that depend on ordinal dissimilarity information only.
\item Combinatorial invariance: outlier prone data that ask for a robust procedure.
\end{itemize}

\subsection{Median and central regions}\label{sec3.2}

Different notions of depth have different medians, which will be discussed next. Also, the shape of central regions and relations to lower-dimensional projections are of interest.

\textbf{Uniqueness of median:}
A median (= point of maximum depth) may be unique or not.
Mahalanobis depth is  uniquely maximized at $E[\bmX]$, zonoid depth too. The halfspace median is unique if $P$ has an L-density and connected support, but generally not; see also Prop.~7 in \cite{MizeraV02}.
Simplicial median is not unique either. The depth $D_{\IL_2}$ takes its {unique} maximum at the {\em spatial median} {\citep{MilasevicD87}}; so does the spatial depth. {Note that in part of the literature this maximum confusably goes under the name \emph{$\IL_1$ median}.}

{Projection median is unique if the location and the spread functionals, here $\med(\bmp'\bmX)$ and $\mad(\bmp'\bmX)$, are continuous as functions of
$\bmp\in S^{d-1}$ \citep{Zuo13}.}

Onion depth is maximum at the innermost convex contour, and this maximum is generally non-unique. Under central symmetry the onion depth is obviously maximal at the center (but not under angular or halfspace symmetry).

Oja depth is maximum at the Oja median, which is not
unique.
It minimizes the average volume of simplices
\begin{eqnarray*}
&& {n \choose d}^{-1} \sum_{1\le i_1<\dots <i_{d}\le n}
\vol_d(\conv \{ \bmy, \bmx_1,\ldots ,\bmx_{d}\})\\[2mm]
&&=\frac {(n-d)!}{n!} \sum_{1\le i_1<\dots <i_{d}\le n}
|\mbox{det}((1,\bmy^\top)^\top,(1,\bmx_{i_1}^\top)^\top,\ldots,(1,\bmx_{i_d}^\top)^\top)|\, .
\end{eqnarray*}

\textbf{Convex or star-shaped level sets:}
  Most depths have convex central regions, that is, are \emph{convex unimodal} functions about their median. Among them are the Mahalanobis, $\IL_p$, halfspace, simplicial volume and zonoid depths. On L-continuous angular-symmetric distributions, simplicial depth is  \emph{star unimodal} about the median, having starshaped central regions \citep{Liu90}.
  The spatial depth does not satisfy monotonicity on rays; so it has neither convex nor starshaped level sets \citep{Nagy17}.

\textbf{Projection property:}
  Some depths have the \emph{projection property},
  \begin{equation}\label{projinf1}
D(\bmy|P) = \inf_{\bmp\in S^{d-1}}D(\langle \bmp,\bmy \rangle|P_\bmp) \, , \quad \bmy \in \IR^d \,,
  \end{equation}
where $\bmX \sim P$ and $P_{\bmp}$ is the distribution of the random variable $\langle \bmp, \bmX \rangle$, that is, of $\bmX$ projected on a ray from $\bmath0$ in direction $\bmp$. A depth that satisfies (\ref{projinf1}) possesses convex level sets.

For example, the halfspace depth and the projection depth satisfy the projection property, as well as the zonoid depth and the Mahalanobis depth; see \cite{Dyckerhoff04}. The projection property  (\ref{projinf1}) allows to approximate a depth value from above by evaluating the univariate depth of projected data at a finite number of directions $\bmp$ and taking the minimum; see Section \ref{sec5.2}.

\subsection{Uniqueness and continuity}\label{sec3.3}

If a depth serves as part of a more complex statistical procedure,
one may be interested in properties of it beyond empirical distributions.

\textbf{Population version:}
Halfspace and simplicial depth have a population version for general distributions, while zonoid and Mahalanobis depth extend to distributions with finite first resp. second moments.
Mahalanobis depth, being a continuous function of moments, obviously satisfies a Law of Large Numbers; the same holds for halfspace depth \citep{DonohoG92}, simplicial depth \citep{Duembgen92}, and zonoid depth \citep[Th.~4.6]{Mosler02a}.
However, convex hull peeling depth, being popular in data analysis, is restricted to empirical distributions.

\textbf{Information on $P$:}
Another important feature of a depth is how much information it carries about the underlying distribution $P$, that is,
how far $P$ is identified, given $D(\bmy|P)$ for all $\bmy$.
While the Mahalanobis depth identifies the first two moments of $P$ only,  the zonoid depth fully determines $P$.
  Halfspace depth identifies the distribution uniquely if the distribution is finite discrete \citep{Koshevoy02}, while in general it does not for infinite discrete or continuous distributions;
  see \cite[Th.~34]{NagySW19} and \cite{Nagy20}.
The Oja depth determines the distribution uniquely among those measures
which have compact support of full dimension \citep{Koshevoy03}. With
  simplicial depth holds the same for empirical distributions in general position \citep{Koshevoy97}.
  If the analysis is restricted to a family of elliptically symmetric distributions having a common strictly monotone decreasing radial density, a distribution is uniquely determined by
  any depth $D$ that is  affine-invariant and continuous in $\bmy$.

\textbf{Continuity on $\bmy$ and $P$:}
For numerical calculations it is important, that the depth depends continuously on the data, i.e., that $D$ be continuous as a function of $\bmx$ and weakly continuous on $P$, uniformly in $\bmx$. More precisely, for some class of distributions $\cal P$ may hold:
\begin{align}\label{continuousx}
\lim_{\bmy_n \to \bmy} & |D(\bmy_n|P) - D(\bmy|P)| =0 \quad \\ & \text{for any sequence}\; (\bmy_n) \;\text{converging to} \;\bmy\,, \nonumber
\end{align}
\begin{align}\label{continuousP}
\lim_{n\to\infty} \sup_{\bmy\in C} & |D(\bmy|P_n) - D(\bmy|P)| = 0 \quad \\ & \text{for any weakly converging $P_n \Rightarrow P$ in $\cal P$}\,. \nonumber
\end{align}

\begin{landscape}
\begin{table}[t!]
\caption{Principal analytical properties of depth notions; $d\ge 2$. Y: Yes; N: No. Ya: Yes for angular symmetric $P$; Ys: for centrally symmetric $P$: Yc: for L-continuous $P$; Yd: for discrete $P$; Ye: for empirical $P$;  Yw: after whitening; Y$_{p=2}$: for $p=2$. Regularity conditions like compact or full-dimensional convex support, uniform integrability of distributions, general position of data, etc are omitted.}\label{table2}
{
\begin{center}
  \begin{tabular}{|l|l|l|l|l|l|l|l|l|l|l|l|}
  \hline
  Property & $D_{\rm {Mah}}$ & $D_{\rm {\IL_p}}$ & $D_{\rm H}$ & $D_{\rm {Proj}}$ & $D_{\rm {Sim}}$ & $D_{\rm {Oja}}$ & $D_{\rm {Zon}}$ & $D_{\rm {Spa}}$ & $D_{\rm {Lens}}$ &  $D_{\rm {Skel\beta}}$ & $D_{\rm{Onion}}$ \\
  \hline
  invariant w.r.t.: &&&&&&&&&&&  \\
  ${\cal T}_{T}\cup {\cal T}_{O}\cup {\cal T}_{uSc}$ 
  & Y & Y$_{p=2}$ & Y & Y & Y & Y & Y & Y & Y & Y & Y \\
  ${\cal T}_{A}$ 
  & Y & Y$_{p=2}$w & Y & Y & Y & Yw & Y & Yw & Yw & Yw & Y \\
  ${\cal T}_{Com}$ 
  & N & N & Y & N & Y & N & N & N & N & N & Y \\
  \hline
max at point of:  &&&&&&&&&&&\\
  central symmetry & Y & Y & Y & Y & Yc & Y & Y & Y & Y & Y & Y \\
  angular symmetry & N & Y$_{p=2}$w & Y & Y & Yc & N & N & Y & N & N & N \\
  halfsp.\ symmetry & N & Y$_{p=2}$w & Y & Y & Yc & N & N & N & N & N & N \\
  \hline
  unique median & Y & Y  & Yc & Y & Yc & N & Y & Y & N & N & N\\
  convex regions & Y & Y & Y & Y & N & Y & Y & N & N & N & Y \\
  starshaped regions & Y & Y & Y & Y & Yca & Y & Y & N & Ycs & Ycs & Y \\
  \hline
  population version & Y & Y & Y & Y & Y & Y & Y & Y & Y & Y & N \\
  continuous on $\bmy$ & Y & Y & Yc & Y & Yc & Y & Y & Y & Yc & Yc & N \\
  continuous on $P$ & Y & Y & Yc & Y & Yc & Ye & Y & Y & Y & Y & N \\
  \hline
  uniqueness of $P$ & N &   & Yd &  & Ye & Y & Y & Y &  &  & N \\
  projection prop. & Y &  & Y & Y & N &  & Y & N & N & N & \\
  \hline
  \end{tabular}
\end{center}
}
\end{table}
\end{landscape}

Mahalanobis depth is continuous (\ref{continuousx}) on $\bmy\in \IR^d$ and meets (\ref{continuousP}) on distributions having a regular covariance matrix. Under a slight regularity condition \citep{CascosLD16}, zonoid depth satisfies (\ref{continuousx}) at every $\bmy\in \IR^d$ and (\ref{continuousP}) on distributions having finite first moment. The same holds for $\IL_2$ depth.
Halfspace and simplicial  depths are in general non-continuous,
but depend continuously on $\bmy$ and $P$ if $P$ has an L-density;
see \cite[Th.~2 and 5]{Liu90} and \cite[Prop.~1]{MizeraV02}.
Oja depth is obviously continuous on $\bmy$ as well as on an empirical distribution $P$.

Table~\ref{table2} summarizes the principal properties of the depth notions considered.

\section{Specifics of the data}\label{sec4}

Special aspects of the data can also guide in selecting a particular depth notion.
E.g., some properties of the data generation process may be known from the setting of the task, like symmetries or the proneness of the process to produce outliers. Further, parts of the data may be missing or {distances may be only ordinarily scaled}.

\subsection{Symmetry}\label{sec4.1}
In many applications {some qualitative} prior information on symmetries of the data generating process {may be available yielding} data that is close to being point symmetric, angular symmetric, elliptically or spherically symmetric.

{If a distribution is elliptical}, the central regions of every affine invariant depth will be {ellipsoids coinciding} with the density level sets.
Under this assumption, Mahalanobis depth is completely satisfactory and also fast to {compute with data}.

\begin{figure}[!t]
\begin{tabular}{ccc}
\,\,\,\,\,Mahalanobis, $n=100$ & Tukey, $n=100$ & Onion, $n=100$ \\
	\includegraphics[width=0.31\textwidth,trim=15 30 15 42.5,clip=true,page=1]{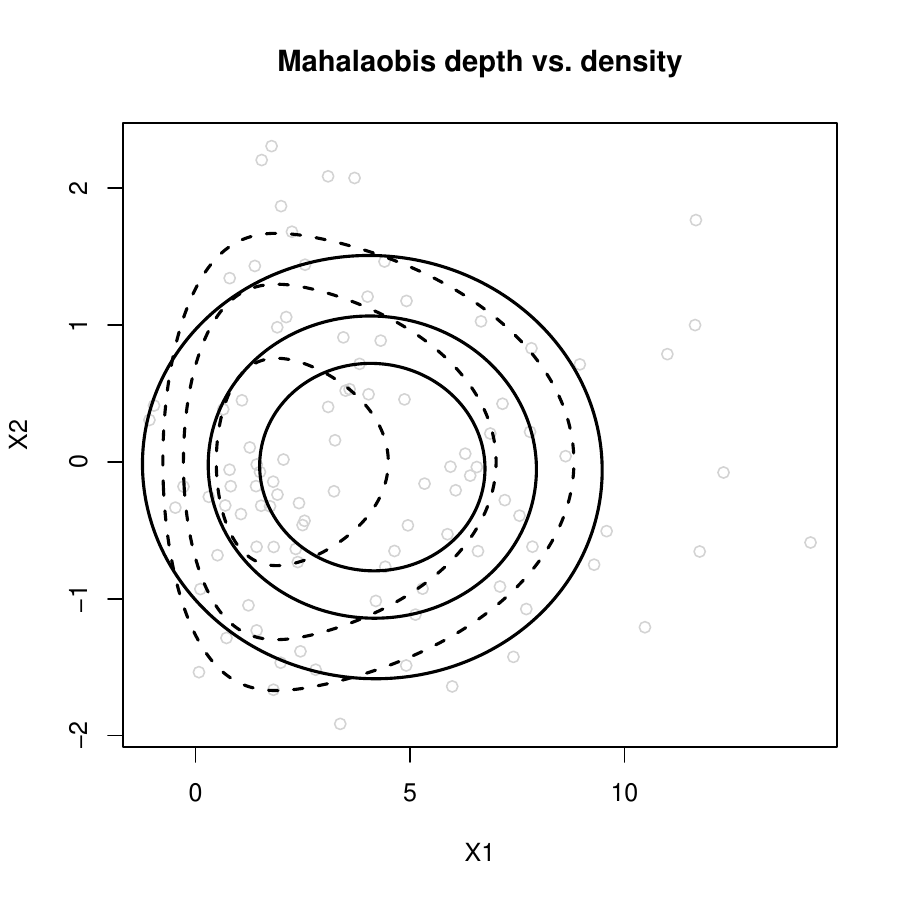} & \includegraphics[width=0.31\textwidth,trim=15 30 15 42.5,clip=true,page=2]{pic-skewed_n100.pdf} & \includegraphics[width=0.31\textwidth,trim=15 30 15 42.5,clip=true,page=3]{pic-skewed_n100.pdf} \\
\,\,\,\,\,Mahalanobis, $n=1\,000$ & Tukey, $n=1\,000$ & Onion, $n=1\,000$ \\
	\includegraphics[width=0.31\textwidth,trim=15 30 15 42.5,clip=true,page=1]{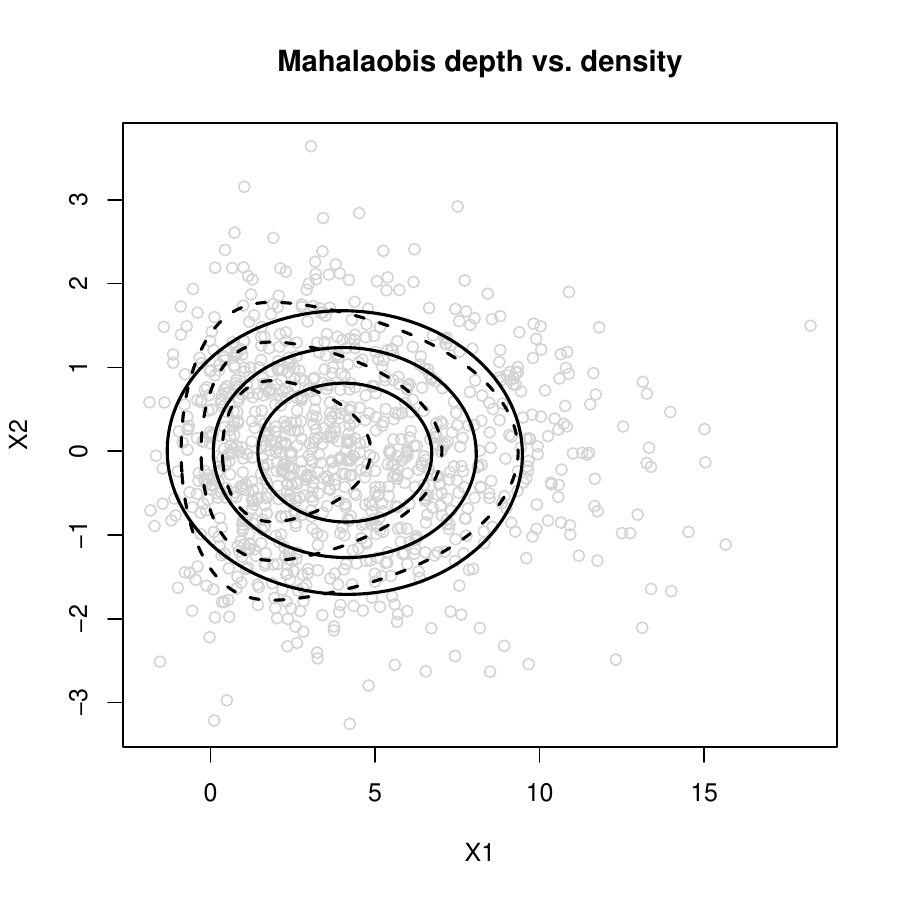} & \includegraphics[width=0.31\textwidth,trim=15 30 15 42.5,clip=true,page=2]{pic-skewed_n1000.pdf} & \includegraphics[width=0.31\textwidth,trim=15 30 15 42.5,clip=true,page=3]{pic-skewed_n1000.pdf} \\
\,\,\,Mahalanobis, $n=10\,000$ & Tukey, $n=10\,000$ & Onion, $n=10\,000$ \\
	\includegraphics[width=0.31\textwidth,trim=15 30 15 42.5,clip=true,page=1]{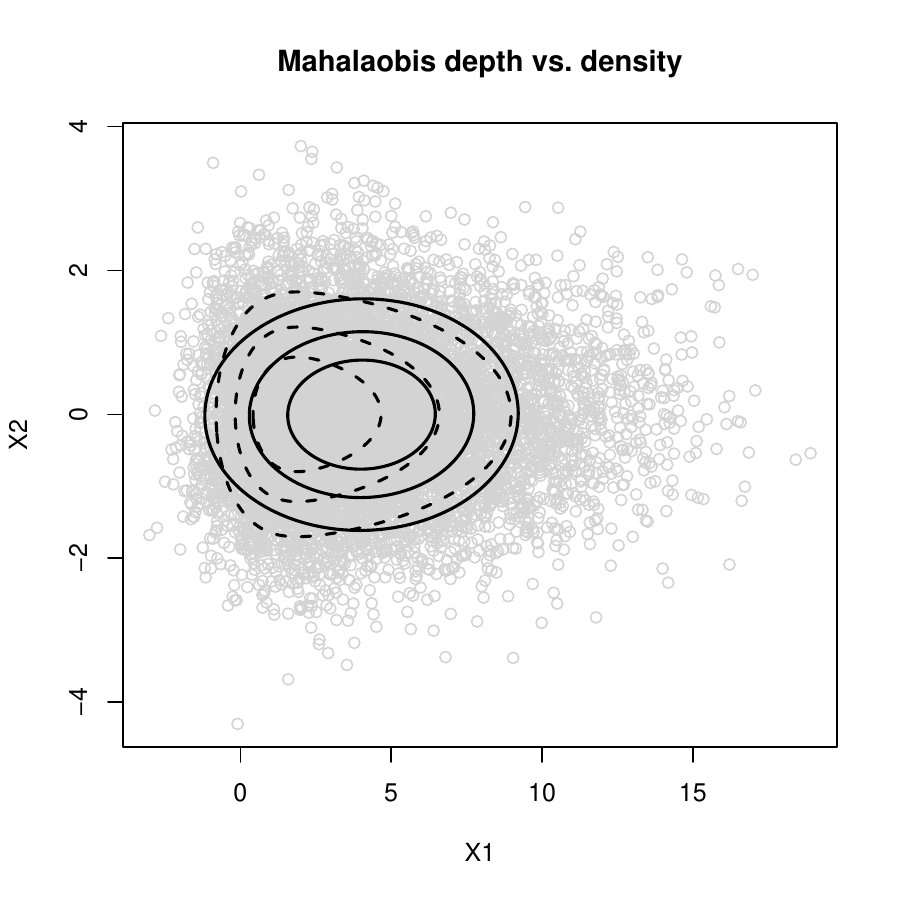} & \includegraphics[width=0.31\textwidth,trim=15 30 15 42.5,clip=true,page=2]{pic-skewed_n10000.pdf} & \includegraphics[width=0.31\textwidth,trim=15 30 15 42.5,clip=true,page=3]{pic-skewed_n10000.pdf}
\end{tabular}
      \caption{Comparison of three depth (solid lines) and density (dotted lines) contours for a bivariate skewed normal distribution. The contours encompass $0.75$, $0.5$ and $0.25$ deepest resp. densest points. Mahalanobis (left column), Tukey (middle column) and onion (right column) depths for a sample of $n=100$ (top row), $1\,000$ (middle row) and $10\,000$ (bottom row) points.}
      \label{fig:skewed}
\end{figure}

But observe that in general depth and density are genuinely different concepts. This is obvious for Mahalanobis regions when the the distribution is non-elliptic.
For illustration, see Figure~\ref{fig:skewed}. It exhibits depth central regions and density level sets having the same probability content and being generated by
a skewed bivariate Gaussian law \citep{AzzaliniDV96},
for Mahalanobis, Tukey and onion depth, and sample sizes $n=100, 1000, 10000$.

Also a possible proneness of the data generation process to producing outliers may be typical in a given task. For robustness, see the next Section~\ref{sec4.2}.

\subsection{Possible outliers}\label{sec4.2}
If the data is suspected to be contaminated by outliers, one may be interested in a robust procedure that reduces the influence of outliers in an automatic, built-in way. Several notions of depth are known to be more or less suitable to downsize the weight of possibly outlying data.

\begin{figure}[!h]
\begin{center}
\begin{tabular}{ccc}
\quad {\small Robust affine invariant} & \quad {\small } & \quad {\small } \\
\quad {\small spatial depth} & \quad\,\,{\small Robust Mahalanobis depth} & \quad {\small Mahalanobis depth} \\
\includegraphics[width=0.31\textwidth,trim = 0cm 0.5cm 1cm 1.5cm,clip=true,page=14]{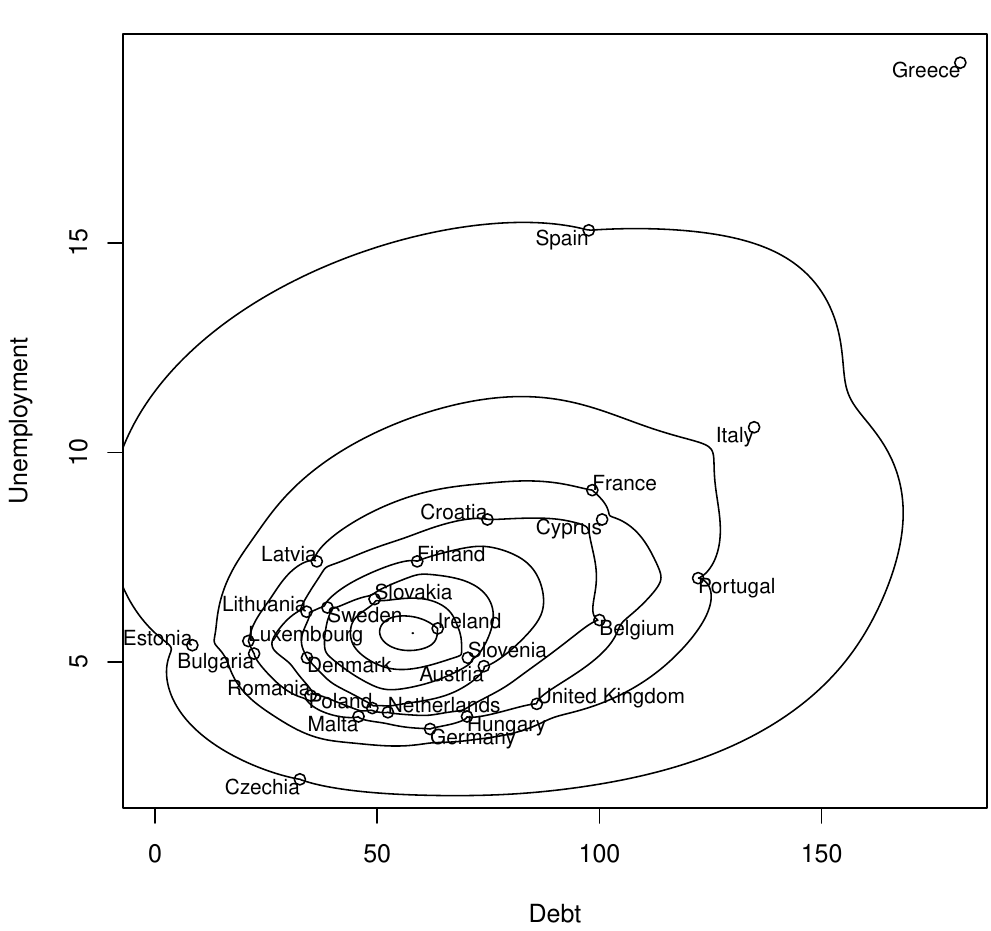} & \includegraphics[width=0.31\textwidth,trim = 0cm 0.5cm 1cm 1.5cm,clip=true,page=3]{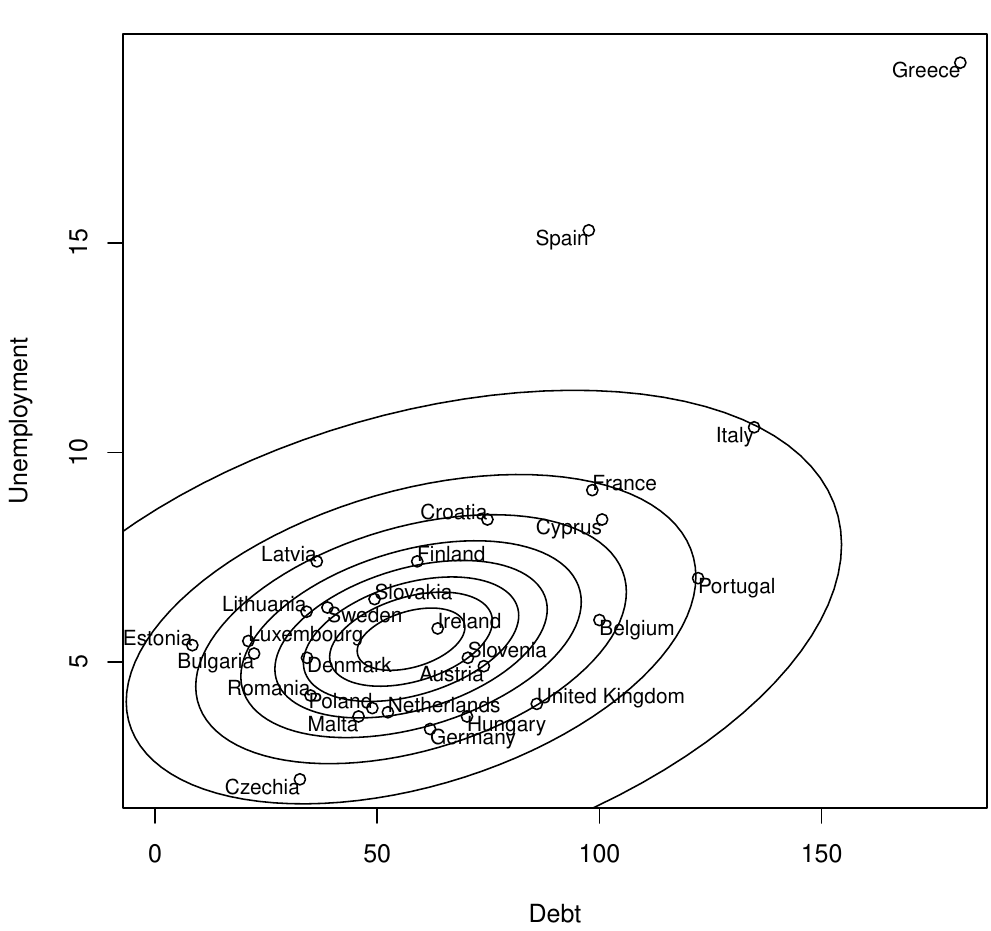} & \includegraphics[width=0.31\textwidth,trim = 0cm 0.5cm 1cm 1.5cm,clip=true,page=2]{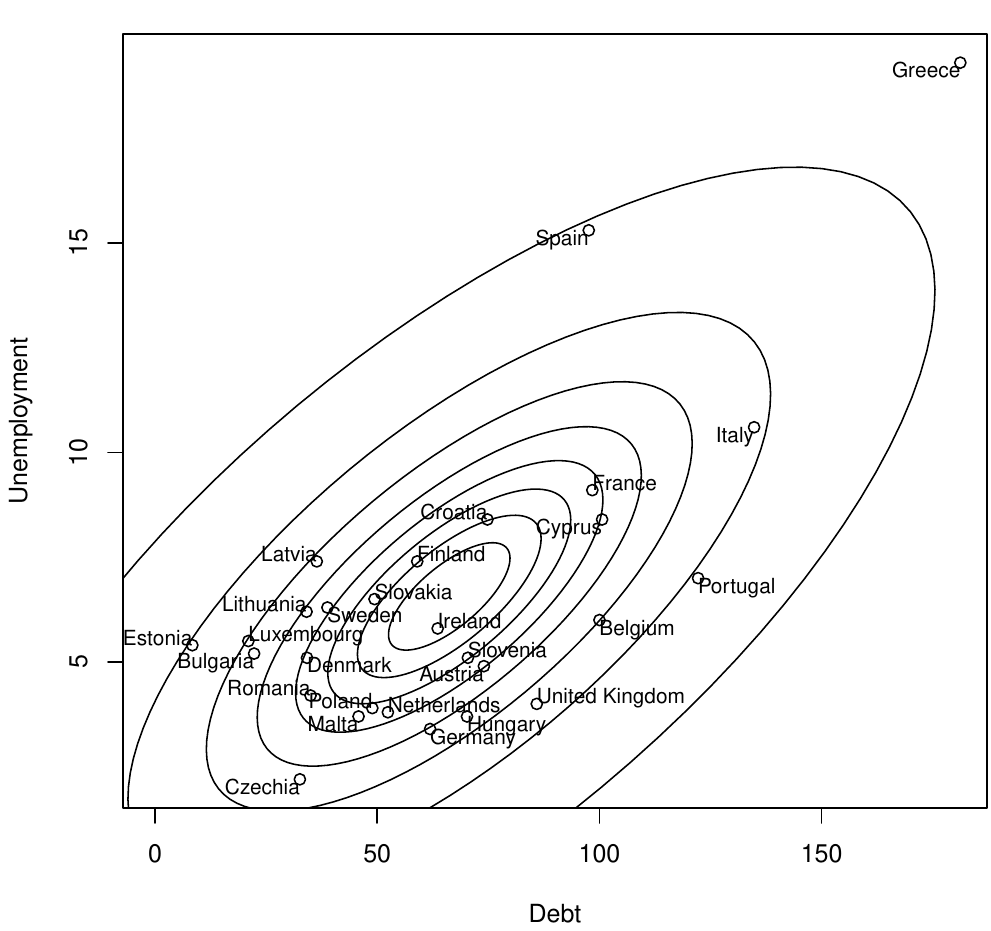} \\
\quad {\small Robust affine invariant} & \quad {\small Robust affine invariant} & \quad {\small Robust affine invariant} \\
\quad {\small simplicial volume depth} & \quad {\small $\IL_2$ depth} & \quad {\small lens depth} \\
\includegraphics[width=0.31\textwidth,trim = 0cm 0.5cm 1cm 1.5cm,clip=true,page=16]{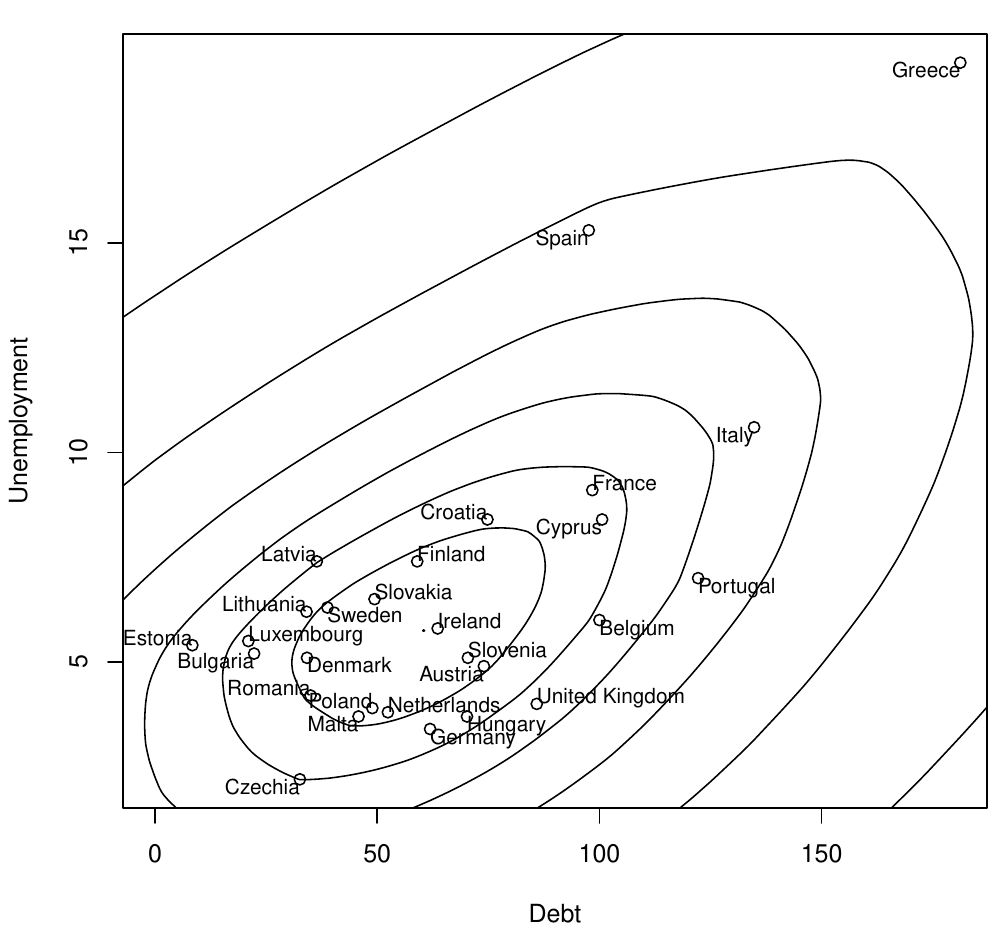} & \includegraphics[width=0.31\textwidth,trim = 0cm 0.5cm 1cm 1.5cm,clip=true,page=20]{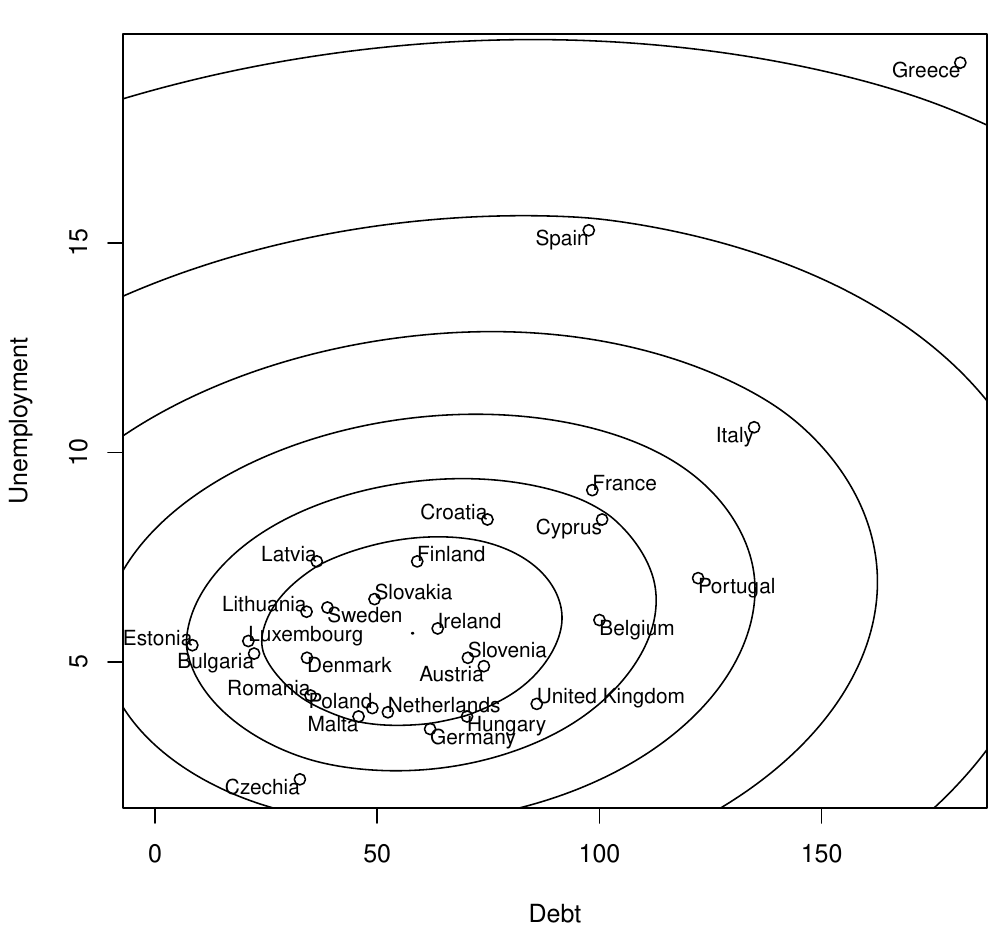} & \includegraphics[width=0.31\textwidth,trim = 0cm 0.5cm 1cm 1.5cm,clip=true,page=18]{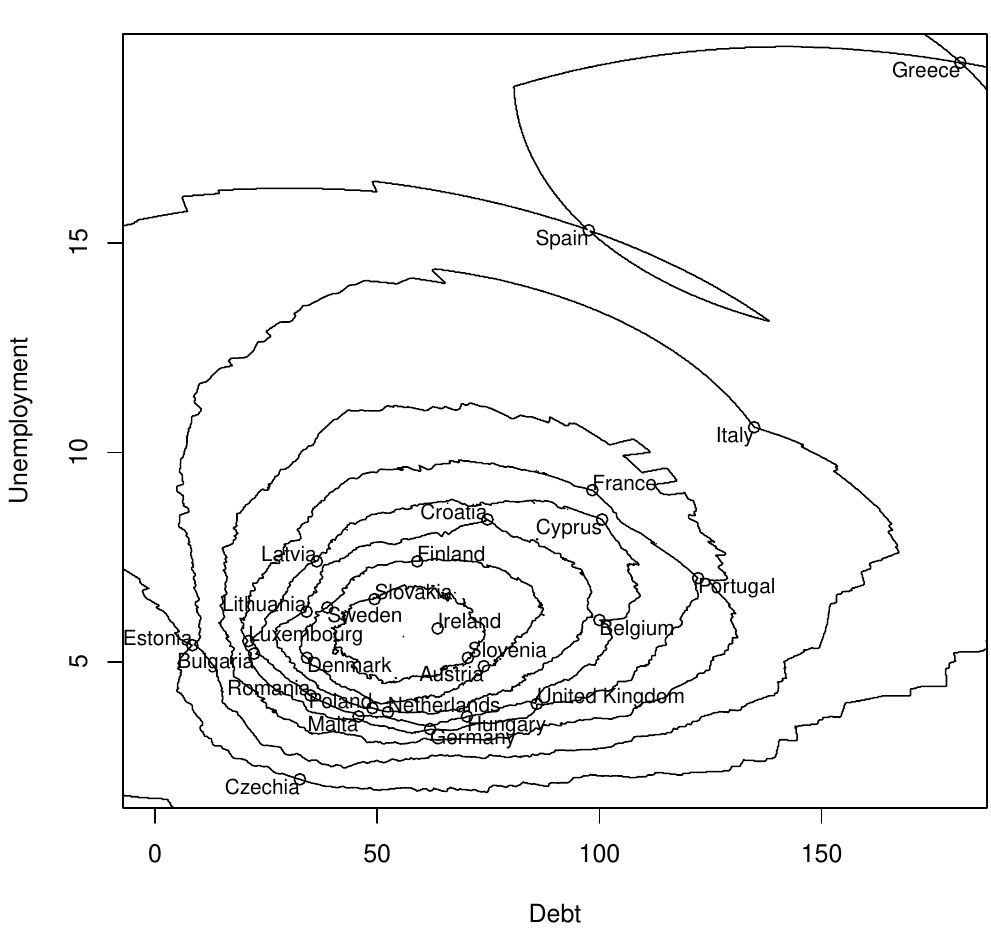} \\
\end{tabular}
\end{center}
\caption{Central regions of EU data on unemployment and public debt in 2018 (Source: EUROSTAT): Five notions of depth, made affine invariant by using a robust scatter matrix (MCD with parameter $\alpha=.75$); for comparison (upper right panel), the non-robust Mahalanobis depth, being calculated with the usual covariance matrix.}\label{fig3}
\end{figure}

{If outliers are expected} to occur, a robust depth notion should be used. A combinatorial depth, like halfspace or simplicial depth, is robust but comes at considerable computational cost. If ellipticity (except outliers, of course) is assumed, the Mahalanobis depth with a robust covariance estimate (e.g. MCD) {may be employed, which is computationally cheaper}.
Spatial depth is robust as well, and fast to compute, but not affine invariant, which can be healed by a robust scatter transformation, see Section~\ref{sec3.1} above.
Figure~\ref{fig3} exhibits central regions of several depths when the data is subject to an MCD scatter transformation.
However, note that plugging-in a robust scatter estimator in place of the usual covariance matrix can influence the stochastic properties of the depth statistic; see \cite{NordhausenT15}.


An indicator of the robustness of a depth is the asymptotic breakdown point of its median. It forms an upper bound of the breakdown of any depth region. The breakdown of the Tukey median is at least $1/(d+1)$ \citep[Prop. 3.4]{DonohoG92}. The Oja median possesses breakdown $0$ \citep{NiinimaaOT90}, hence all simplicial volume regions have breakdown $0$. The same holds for onion regions \citep[Sec.\ 4]{DonohoG92}. Also, zonoid regions and, more general, weighted mean regions have breakdown $0$ as their median equals the mean of the data. The breakdown of the $\IL_2$-depth median ( = spatial median = $\IL_1$ median) is $1/2$ \citep{LopuhaaR91}.
However the breakdown of the depth function $D_{\rm {\IL_p}}(\bmy)$ at some point $\bmy$
can be much higher \citep{Zuo04}.
Another indicator of robustness of a depth is the influence function of its median; it is investigated in \cite{Romanazzi01} for halfspace depth, in \cite{NiinimaaO95}  for $\IL_2$, spatial, and Oja depth, in \cite{Zuo04} for $\IL_p$ depth, and in \cite{Zuo06} for projection depth.

Also, the direction of outliers can be relevant.
 If one is interested in the direction of an outlier, a non-robust depth may be employed which, like the zonoid depth, possesses the projection property. Then a point is identified as an outlier, if it has low depth value, minimized over the relevant directions.

\subsection{Missing data}
In principle, a data depth $D(\bmy|\bmX)$ is evaluated at the point $\bmy$ and $n$ points representing the support of $\bmX$, which data is usually given as a complete $(n+1)\times d$-matrix of real numbers. Functional data may be treated, after proper discretization, like multivariate data; see also Section \ref{sec6.2} below.
However, some depth notions can cope with incomplete data.

When coordinate values of $\bmy$ are missing, the depth calculation may be restricted to a lower-dimensional space of attributes, corresponding to the values that are non-missing. If the depth satisfies the projection property (\ref{projinf1}), this yields an upper bound of the unknown depth value.

When coordinate values are missing of points in the support of $\bmX$, a lower bound on zonoid depth $D_{\rm {Zon}}(\bmy|\bmX)$ can be determined as follows. Let values of the $j^*$-th coordinate, $x_{ij^*}$, be missing at observation units $i\in J\subset \{1,2,\dots, n\}$. If, as usual, for each of the missing values the arithmetic mean $\overline{x}_{j^*}=\ave_{i\not\in J} x_{ij^*}$ is imputed, a lower bound of the depth is achieved. To see this, note that in Formula (\ref{ZonoidDataDepth}) the restriction
\[ y_{j^*} = \sum_{i=1}^n \lambda_i x_{ij^*}\,, \quad \text{that is} \quad y_{j^*} - \overline{x}_{j^*}= \sum_{i=1}^n \lambda_i (x_{ij^*}-\overline{x}_{j^*})\,,
\]
arises. By the imputation, the latter summands with $i\in J$ will become zero, so that at least one of the weights $\lambda_i$ will decrease (in the weak sense), hence
$D_{\rm {Zon}}(\bmy|\bmX)= \inf_i (n \lambda_i)^{-1}$ not be lowered.

\section{Computational feasibility}\label{sec5}

For most notions of data depth, the depth of a point can, in principle, be exactly calculated. But
the computational work increases with $n$ and, often exponentially, with $d$. The latter is typical for combinatorial depths.
Then, practical restrictions on computation time and storage space may urge the statistician to rely on approximative approaches. If $n$ is large, thinning the data cloud (by random selection) may solve the task. To cope with a large dimension $d$ comes out to be the much harder problem.
This section surveys exact and approximate procedures to calculate the depth of a point. Also a short account of procedures for calculating depth regions is given. Recent implementations of these procedures are referred to and strategies for large and high dimensional data are discussed.

\subsection{Exact calculations}\label{sec5.1}
Exact procedures have been implemented in the \texttt{R}-package \texttt{ddalpha} to calculate the following nine data depths:
Mahalanobis, $\IL_2$, spatial, zonoid, lens, onion, halfspace, simplicial volume, and simplicial depth.
Note that for most of these depths the naive direct calculation is not feasible, and more sophisticated procedures of lower computational complexity have to be used.

For example, take a look at the halfspace depth. Its definition suggests a simple idea: splitting the sample in two parts and checking whether the parts can be linearly separated by a hyperplane containing $\bmy$. Then, among all partitions separable by a hyperplane, the one is selected that has the smallest number of observations on one side. However, there is a total of $2^n$ possible partitions, which leads to a time complexity exponential in $n$. A closer inspection \citep{DyckerhoffM16} gives an idea how to reduce the set of possible candidates, leading to an algorithm with polynomial complexity in $n$, $O\bigl(n^{d-1}\log(n)\bigr)$.

\begin{figure}[!t]
  \begin{center}
    \includegraphics[keepaspectratio=true,width = \textwidth, trim = 1mm 2mm 0mm 2mm, clip, page = 1]{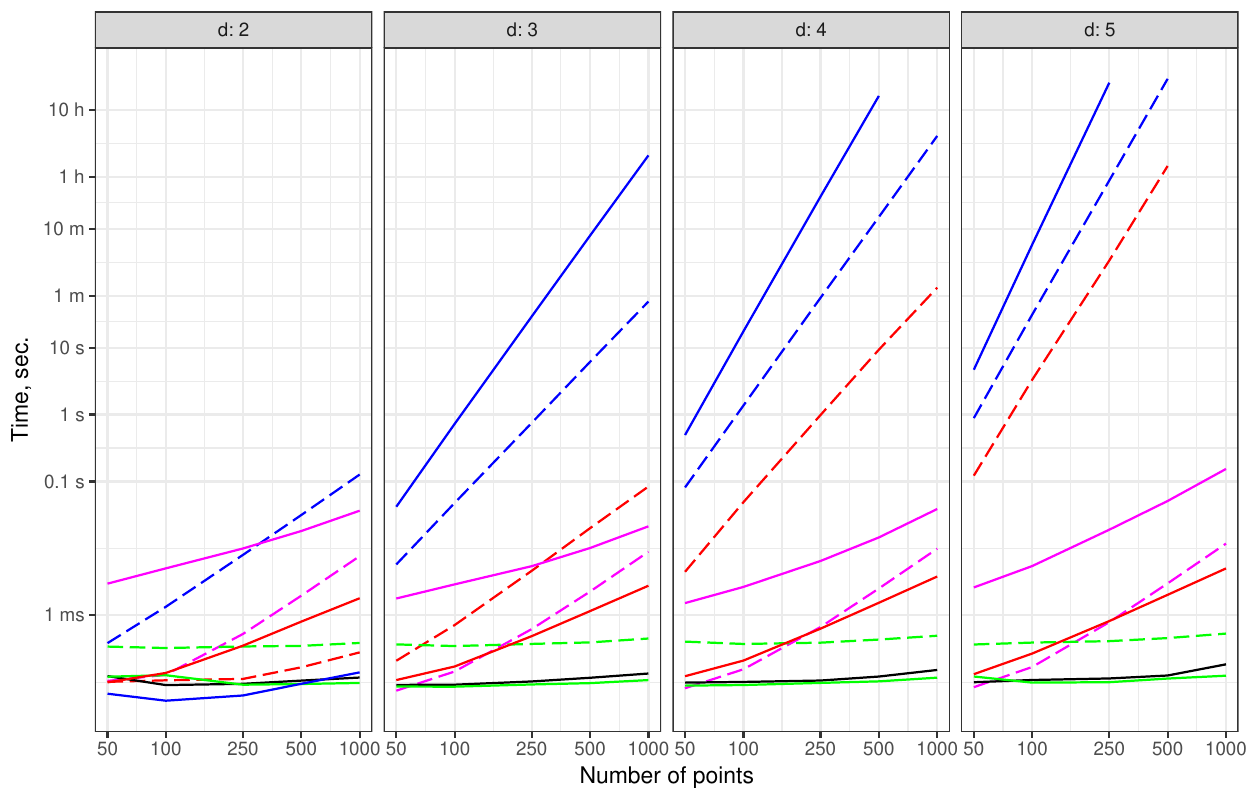}\\
    {\small{\color{red}\LBL} zonoid, {\color{red}\LBD} halfspace, {\color{green}\LBL} Mahalanobis, {\color{green}\LBD} spatial, {\color{black}\LBL} $\IL_2$,\\ {\color{blue}\LBL} simplicial, {\color{blue}\LBD} simplicial volume, {\color{magenta}\LBL} onion, {\color{magenta}\LBD} lens}
    \caption{Calculation time of various depth functions on double logarithmic scale (sample size $n$ and time $t$).}
    \label{fig:depth_speed}
  \end{center}
\end{figure}

Figure~\ref{fig:depth_speed} exhibits average computation times for each of the depths, depending on sample size $n$ and dimension $d$, where $n$ runs up to 1000 and $d= 2,3,4,5$. Given $n$ and $d$, 30 samples have been drawn, depth has been calculated for 25 points of each sample, and an average has been taken over these 25 points and 30 samples.

All calculations have been performed by means of the \texttt{R}-package \texttt{ddalpha} on a machine having processor Intel(R) Core(TM) i7-4980HQ (2.8 GHz) with 16 GB of physical memory and macOS Sierra (Version 10.13.4) operating system.

As is seen from Figure~\ref{fig:depth_speed}, the depth notions differ greatly in their computation times, some sharply increasing with $n$ and $d$. (In each panel, times and sample sizes are measured on a double logarithmic scale.)
Table~\ref{tab:compl} lists the complexities of the various algorithms, including references to the literature.

\begin{landscape}
\begin{table}[t!]
\caption{Time complexities for exact and approximate computation of several depths (of a given point in $\IR^d$ regarding a sample). For approximate computation, $k$ stands for the number of random directions for halfspace, projection and zonoid depth, resp.\ for the number of considered simplices for simplicial and simplicial volume depth, resp.\ for the number of considered pairs of points for $\beta$-skeleton depth.}
	\label{tab:compl}
	\begin{center}
	\begin{tabular}{l|cc}
		Depth notion & Exact & Approximate \\ \hline
		Mahalanobis & $O(n)$ & --- \\
		$\IL_p$ & $O(n)$ & --- \\
		halfspace & $O\bigl(n^{d-1}\log(n)\bigr)$, $O(n^d)$ & $O(kn)$ \\
		& \cite{RousseeuwS98} & \cite{Dyckerhoff04} \\
		& \cite{DyckerhoffM16} & \cite{MozharovskyiML15}  \\
		projection & $O(n^d)$, \cite{LiuZ14} & $O(kn)$, \cite{Dyckerhoff04} \\
		simplicial & $O(n^{d+1})$ & $O(k)$ number-approx. \\
		& & $O(n^{d+1})$ portion-approx. \\
		simplicial volume & $O(n^{d})$ & $O(k)$ number-approx. \\
		& & $O(n^{d})$ portion-approx. \\
		zonoid & \cite{DyckerhoffKM96} & $O(kn)$, \cite{Dyckerhoff04} \\
		spatial & $O(n)$ & --- \\
		$\beta$-skeleton & $O(n^2)$ & $O(k)$ \\
		onion & $O(n^{\lfloor d/2 \rfloor}/d^d)$, \cite{BarberDH96} & --- \\
	\end{tabular}
	\end{center}
\end{table}
\end{landscape}

\begin{itemize}
  \item Mahalanobis depth and $\IL_2$ depth, as well as spatial depth, are always quickly calculated, and their computation times are virtually independent of (moderate values of) $n$ and $d$.
  \item In dimension $d=2$ all depths can be sufficiently fast calculated, with times staying below 100 milliseconds even w.r.t. a sample of $n=1000$ points.
  \item When $d\ge 3$, the two combinatorial invariant depths, simplicial and halfspace depth, need much more computation time $t$, and at given dimension this time grows with $n$ as $t=n^{const_d}$.
      The increase in $d$ is exponential, see Table~\ref{tab:compl} regarding computational complexity.
      The same holds for the simplicial volume depth. Of these three depth notions, the simplicial depth is uniformly slowest, while the halfspace depth outperforms the other two. This experimental result corresponds to the respective complexities of order $n^{d+1}$, $n^d$, $n^{d-1}\log(n)$; see Table~\ref{tab:compl}.
  \item The time needed for the zonoid depth grows much slower with $n$ and only slightly with $d$.
  \item The onion depth needs always more time than the zonoid depth, but has similar growth behavior.
\end{itemize}

Mostly, there is a trade-off between the accuracy of a procedure and its speed.
From Table~\ref{tab:compl} is seen that Mahalanobis, $\IL_p$, and spatial depth
have the best possible time complexity $O(n)$ (as one cannot consider all points without looking at least once at each of them).
Also, exact algorithms for zonoid (linear programming, which is known to be usually efficient) and $\beta$-skeleton depths are sufficiently fast; see Figure~\ref{fig:depth_speed}.
A rather paradoxical result appears,
when the Tukey depth of all points of a given sample is calculated with respect to the same sample. In this case, the complexity of computing the depth of a single point is lower than linear in $n$, \textit{viz.} $O(k \log(n))$. For details see Section~2.3 of \cite{MozharovskyiML15}.

\subsection{Approximate calculations}\label{sec5.2}

Clearly, if an exact procedure is available and time and memory space allow, the depth of a point should be calculated by the exact procedure. This is the obvious ``gold standard''. However, it may be non-feasible in practice if $d$ and/or $n$ are too large or if the depth has to be very often evaluated
\begin{itemize}
  \item  as in bootstrap or permutation procedures, or
  \item  when a whole central region is calculated.
\end{itemize}

Simplicial and simplicial volume depths can be approximated by considering either a fixed number or a constant portion of all simplices; in the latter case the complexity amounts to that of the exact algorithm, though. The respective complexities are exhibited in Table~\ref{tab:compl}. To determine the onion depth in higher dimensions, the convex hull can often only be approximately calculated. This may yield
ambiguous results and affect the computation of the onion depth.

{If a depth satisfies the projection property (\ref{projinf1}) it may be approximated by using finitely many directions
and taking the minimum, which yields an upper bound of the depth.

Not every depth satisfies the projection property, and thus can be approximated with random directions. Some depths have non-convex  regions, like the spatial and the simplicial depth. Some need drawing simplices, like the Oja depth and the simplicial depth.
The approximation from above may, in particular, be worthwhile for the halfspace depth and the projection depth.
(Recall that Mahalanobis and zonoid depth can be exactly calculated at low cost.)

As the halfspace (= Tukey) depth $D_{\rm H}$ satisfies the projection property, an upper bound of the depth is obtained by
using a finite set $S\subset S^{d-1}$ of directions and taking the minimum,
  \begin{equation}
D_{\rm{RTD}}(\bmy|P) = \min_{\bmp\in S}D_{\rm H}(\langle \bmp,\bmy \rangle|P_\bmp) \,.
  \end{equation}
If information about `favourable' directions is known, these directions should be primarily pursued. If not, directions are randomly chosen on the unit sphere, which is known as the \textit{random Tukey depth} \citep{CuestaAN08b}.
However, the number of directions needed for a given precision is not known; for some results on convergence rates, see \cite{NagyDM19}.

Figure~\ref{figure:RTDexact1} shows how often an approximation by the random Tukey depth (RTD) hits the correct value of the Tukey depth if just 1000 random directions are chosen. While in dimension $d=2$ this is virtually always the case, in dimension $d=3$ (resp.\ $d=4$ and $d=5$) the portion of correct values decreases with the sample size $n$, attaining less than 50 percent (resp.\ 25 percent) at $n=300$. Presumably, the number of needed directions  goes exponentially with $d$; see also \cite{CuestaAN08b}.

\begin{figure}[!h]
  \begin{center}
    \includegraphics[keepaspectratio=true,width = \textwidth, trim = 1mm 2mm 0mm 2mm, clip, page = 1]{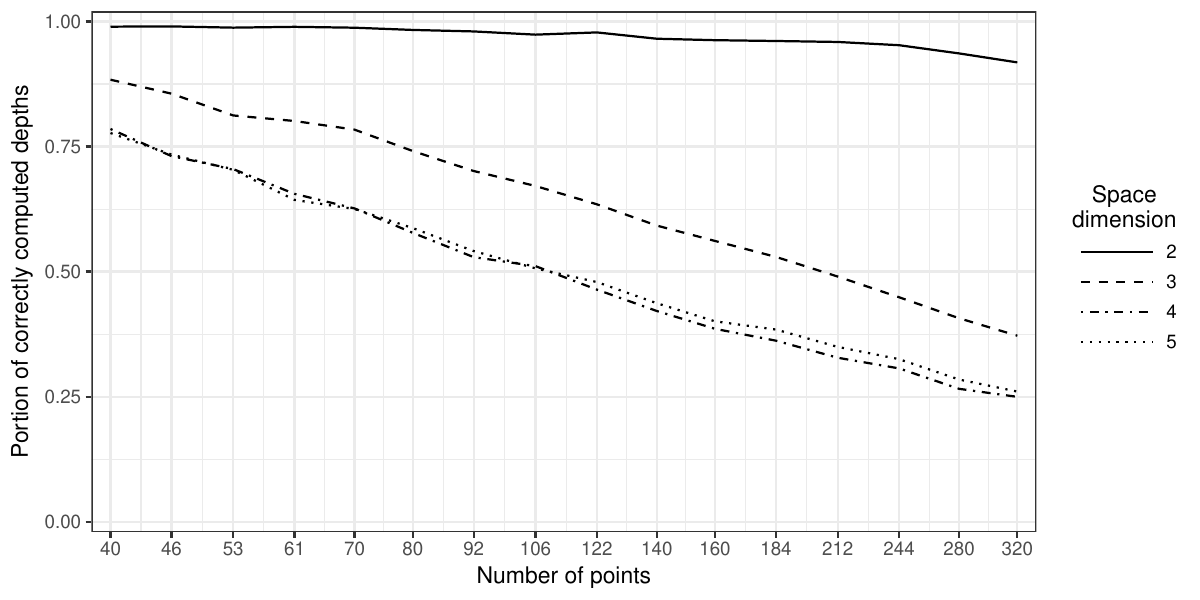}
    \caption{Average rates of achieving the exact value of Tukey depth when calculating the random Tukey depth at
    $1000$ directions. The directions are uniformly distributed on the unit sphere, and the average is taken over all points of a standard normal sample with respect to the remaining sample, given $d$ and $n$.}\label{figure:RTDexact1}
  \end{center}
\end{figure}

Of interest is the \textit{absolute error} incurred by this approximation, $ D_{\rm{RTD}}(\bmy|P) - D_{\rm H}(\bmy|P)$, as well as
the \textit{relative error}, ${D_{\rm{RTD}}(\bmy|P) - D_{\rm H}(\bmy|P)}/{D_{\rm H}(\bmy|P)}$.

Figures~\ref{figure:RTDabsolute} and \ref{figure:RTDrelative} exhibit boxplots of relative and absolute errors of the random Tukey depth for normally distributed data, when a set $S$ of 1000 directions is employed.
As is seen from Figure~\ref{figure:RTDabsolute}, absolute errors stay rather small up to dimension $d=5$. However, Figure~\ref{figure:RTDrelative} reveals that substantial relative errors arise, often amounting to 50\% up to a multiple of the correct value.

  \begin{figure}[h!]
	\begin{center}
    \includegraphics[keepaspectratio=true,width=\textwidth]{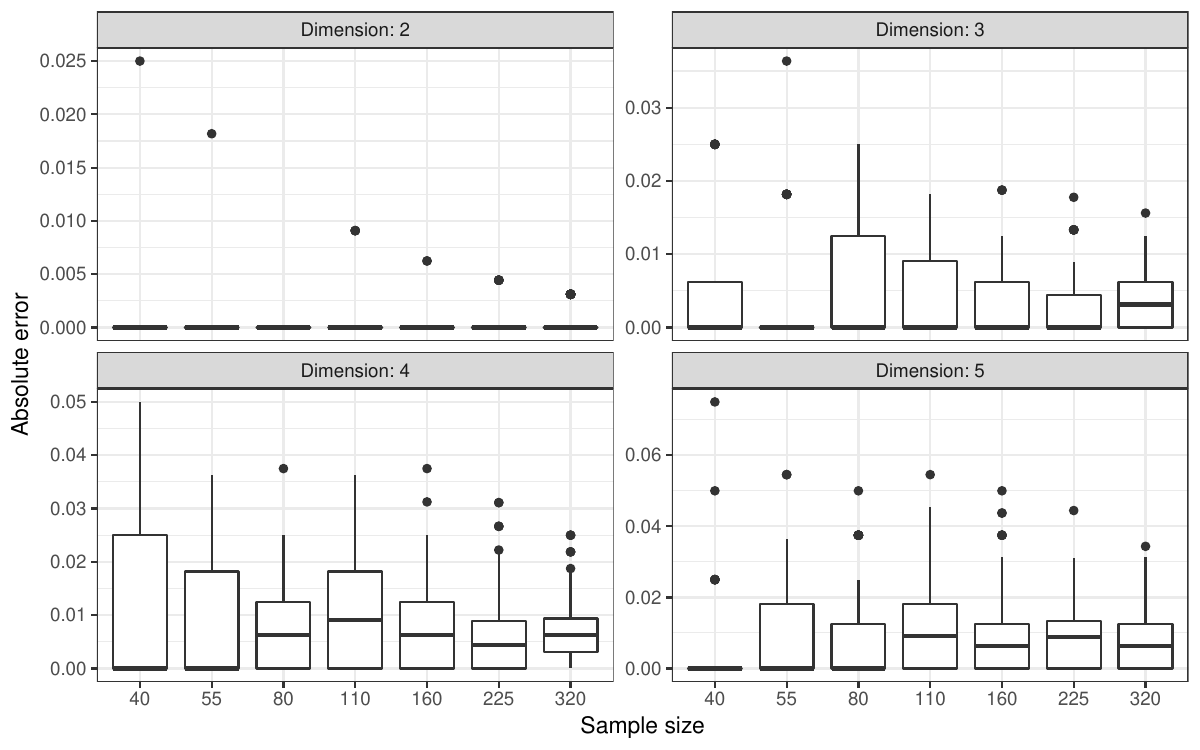}
\caption{Absolute error of the random Tukey depth when approximating at 1000 directions.}\label{figure:RTDabsolute}
  \end{center}
\end{figure}

  \begin{figure}[h!]
	\begin{center}
    \includegraphics[keepaspectratio=true,width=\textwidth]{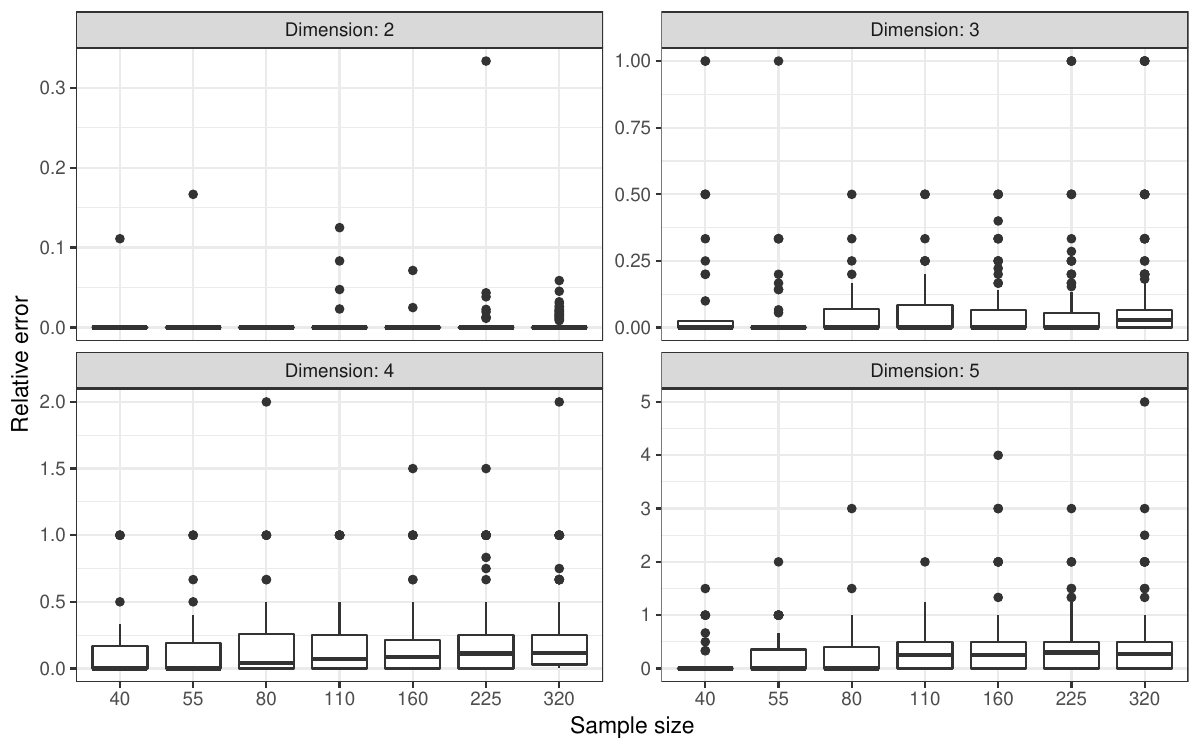}
    \caption{Relative error of the random Tukey depth when approximating at 1000 directions.}\label{figure:RTDrelative}
  \end{center}
\end{figure}

Several other approximate algorithms for Tukey depth have been proposed in the literature; for a comparison see \cite{Zuo19}, who provides a fast approach being able to cope with large $d$.

However, if one is interested in a most precise value of a depth statistic, it is often computationally cheaper  to calculate the exact depth immediately instead of repeatedly seeking for approximate values.

In certain applications many depth values have to be determined, while the precision of each single value is less important.
E.g. in depth-based classification the rule is noisy anyway, so the noise from the approximation may not much influence the results of cross-validation and the finally constructed rule (if at all); see e.g.\ \cite{LangeMM14b}. To give another example, in nonparametric testing based on data depth \citep{Dyckerhoff02}, the Wilcoxon-Mann-Whitney ranks do not very sensibly depend on the single depth values.
In these and many other applications the precision of the depth calculation must be related to the noisiness of the remaining procedure, and more or less rough approximations may be well.

\subsection{Calculation of regions}

Mahalanobis regions are always ellipses; determining them simply requires the calculation of the mean (being the depth median) and the covariance matrix, which has time complexity $O(n)$. When using robust estimates instead, complexity is generally higher.

To compute depth regions in dimension 2, often the idea of a circular sequence \citep{Edelsbrunner87} is exploited.
E.g. \cite{RutsR96} use this sequence in an algorithm for bivariate Tukey regions, which has
complexity $O(n^2\log\,n)$. In a similar way \cite{RousseeuwR98} calculate the bivariate Tukey median with complexity $O(n^2\log^2\,n)$.

For Tukey regions in higher dimensions ($d > 2$) several approaches have been pursued in the literature: Based on \cite{HallinPS10}, \cite{PaindaveineS12a,PaindaveineS12b} construct algorithms for computation of multiple-output quantile regression (yielding Tukey depth contours when regressing on a constant), which have time complexity at most $O(n^{d+1})$, and on an average $O(n^d)$; {see also \cite{HallinLPS15}.} \cite{LiuMM19} suggest a fast algorithm that computes Tukey regions with time complexity $O(n^d\log\,n)$, and also the Tukey median with $O(n^d\log^2\,n)$.
Further references on the computation of Tukey regions are found in these articles.

\cite{BazovkinM12} develop an algorithm to compute general weighted-mean trimmed regions \citep[see][]{DyckerhoffM11}, with complexity varying between $O(d^2n\cdot n^d)$ and $O(d^2n\frac{n^{2d}}{2^d})$ depending on the weighting scheme (which determines the number of facets of the region).
In particular the calculation of zonoid regions, which are special weighted-mean trimmed regions, achieves complexity $O(d^2n\cdot n^d)$.

An algorithm for computation of projection depth contours in any dimension has been constructed by \cite{LiuZ14}, who also compute the projection median. Algorithms for the Oja median are provided in \cite{FischerMMNPV18}, while \cite{KentEC15} \citep[see also][]{Ostresh78} elaborate on the computation of the spatial median. Onion depth contours can be obtained as convex hulls of sequentially peeled data by using the \texttt{qhull} algorithm \citep{BarberDH96}.

\subsection{Implementations}\label{sec5.3}

{Data depths have been implemented in several existing software packages.
In particular,
\texttt{R}-package \texttt{ddalpha} \citep{PokotyloMD19,ddalpha} implements exact procedures for all above mentioned depths, except projection depth. In addition, approximate algorithms are provided for the halfspace, projection, simplicial, and simplicial volume depths. Packages \texttt{depth} \citep{depth}, \texttt{DepthProc} \citep{DepthProc}, \texttt{fda.usc} \citep{FebreroBandeOF12}, \texttt{mrfDepth} \citep{mrfDepth} implement a number of depth notions as well.

Mahalanobis depth is easily coded by hand in any programming language; ready-to-use implementations are also found in \texttt{R}-packages \texttt{DepthProc} and \texttt{fda.usc}.
\texttt{R}-package \texttt{DepthProc} suggests an implementation of $\IL_p$ depth.
Halfspace depth can be computed exactly for $d\le3$ (and approximately) in any dimension with \texttt{R}-packages \texttt{depth} and \texttt{mrfDepth}, and only approximately with \texttt{R}-packages \texttt{DepthProc} and \texttt{fda.usc}. Exact projection depth is computed with \texttt{MATLAB}-package \texttt{CompPD} \citep{LiuZ15}, while approximate procedures are included in \texttt{R}-packages \texttt{DepthProc}, \texttt{fda.usc}, and \texttt{mrfDepth}. Exact simplicial depth for $d=2$ is calculated with \texttt{R}-packages \texttt{depth}, \texttt{fda.usc}, and \texttt{mrfDepth}. Exact simplicial volume depth is also computed using \texttt{R}-package \texttt{depth}. Spatial depth is implemented in \texttt{R}-package \texttt{depth.plot} \citep{depth.plot}.

{Existing software allows for computation of depth regions as well. Tukey regions are calculated with \texttt{R}-packages \texttt{modQR} \citep{modQR} or \texttt{TukeyRegion} \citep[][including the Tukey median]{LiuMM19}  and \texttt{Octave}-package \texttt{modQR} \citep{BocekS16}. Trimmed regions and median of the projection depth are computed using the before mentioned \texttt{MATLAB}-package \texttt{CompPD}. The Oja median is  determined using the \texttt{R}-package \texttt{OjaNP} \citep{FischerMMNPV18}. \texttt{R}-package \texttt{WMTregions} \citep{BazovkinM12} computes zonoid regions \citep[see also][]{MoslerLB09}, as well as trimmed regions for the entire family of weighted-mean depths. Onion depth regions are constructed using the \texttt{qhull} implementation of the \texttt{R}-package \texttt{geometry} \citep{geometry}.}

Obviously, this overview cannot be complete. Moreover, the packages are continuously modified by their authors.}

\subsection{Large and high-dimensioned data}\label{sec5.4}

Depth statistics can also be applied to analyse sets of data having large data size $n$ and/or high dimension $d$.
But the different notions of depth are appropriate to different situations, and sometimes a pre-treatment of the problem may be needed.

Computational feasibility of a depth notion depends on $n$ and $d$ as well as on their relative size.
In most applications $n$ is considerably larger than $d$, $n>>d$.
If not, a large portion of data lies on the border of the data cloud's convex hull and, consequently, has zero depth in all depth statistics that vanish outside this convex hull, \emph{viz.} halfspace, simplicial, zonoid and onion depth, which idles these notions.
If $n<d$ or the sample covariance matrix is ill-conditioned, Mahalanobis depth as well as other whitened depths have to be modified by basing them on another shape matrix.

If both $n$ and  $d$ are large and $n>d$, Mahalanobis depth can routinely be calculated, while for moderate $d$ zonoid, spatial, lens, and  $\IL_p$ depth are computationally feasible.
However if the properties of these depths do not fit to the application problem at hand, $n$ and/or $d$ may be reduced. $n$ is decreased by `thinning', that is selecting a representative part of the sample. To downsize $d$, `features' (= attributes of the sampled items) have to be preselected by recurring to additional information.

\section{Extensions}\label{sec6}

This section shortly presents several extensions of the above.  {Firstly local depth notions are considered, having level sets that are not necessarily nested and starshaped about a common center. Then the Monge-Kantorovich depth \citep{ChernozhukovGHH17} is discussed, which is a
 global depth allowing for non-convex central regions but retaining nestedness and a common center.
Further, depth notions for functional data are mentioned, as many of them are built on multivariate depths. Finally, a more general view on depth as depth of a fit is discussed, which yields extensions beyond locational depth such as regression depth \citep{RousseeuwH99} and tangent depth \citep{Mizera02}.}

\subsection{Local depths}\label{sec6.1}
As observed in Section~\ref{sec4.1}, depth and density are different concepts. The reason is that depth refers to the whole distribution, while density measures it locally. {In particular, depth level sets are starshaped and nested about a common center, while density level sets generally are neither starshaped nor nested but reflect local features such as multiple modes.}

In some instances a kind of depth is asked for which describes local aspects of the distribution. \cite{AgostinelliR11} introduce localized versions of Tukey and simplicial depth. For Tukey depth, they replace the halfspaces in Definition (\ref{Hdepth}) with infinite slabs of finite width $h$, for simplicial depth they restrict to simplices of some given volume $h$. When $h$ goes to infinity, the usual notion is obtained. The smaller $h$, the more local features of the distribution are represented by the depth.
{\cite{HlubinkaKV10} introduce another localized generalization of Tukey depth by imposing a weight function on the halfspaces.
All these depths are orthogonal and translation invariant, but not invariant to scale.}

A depth that is based on point differences, like the spatial depth and the Mahalanobis depth, can be localized as follows: Transform each difference $\bmt$ by a positive definite kernel, $k_h$, e.g. the Gaussian kernel $k_h(\bmt)= (\sqrt{2 \pi h})^{-d}\exp(-||\bmt/h||^2/2)$, and calculate the respective \emph{kernelized depth}. By this approach, \cite{ChenDPB09} introduce
the \emph{kernelized spatial depth}:
\begin{equation}\label{kernelspatial}
   D_{\rm{kSpa}}(\bmy|\bmX) = E[k_h(\bmy - \bmX)] -  \left|\left|E\left[k_h(\bmy - \bmX) \cdot \frac{\bmy-\bmX}{||\bmy-\bmX||}\right]\right|\right|\,.
\end{equation}

$D^*_{\rm{kSpa}}$ is obtained by first whitening $\bmy\cup \bmX$ and then kernelizing the distances.
It, up to scale, approaches density (resp.\ usual spatial depth), when the band width $h$ goes to 0 (resp.\ $\infty$); see Theorem 3 in \cite{DuttaSG16}.
A kernelized Mahalanobis depth is proposed by \cite{HuWWLH11}.

Different from these approaches, \cite{PaindaveineVB13} construct a local depth by conditioning a given (global) depth $D(\bmy|\bmX)$ on a neighborhood of $\bmy$. Instead of the distribution
$P_\bmX$ of $\bmX$, they consider the mixture $P_{\bmy,\bmX} = .5 P_\bmX + .5 P_{2\bmy-\bmX}$, which is a symmetric distribution about $\bmy$. For some probability  $\beta\in]0,1]$, the central region $D_{\rm{\beta}} (P_{\bmy,\bmX})$
serves as a local neighborhood of $\bmy$. Conditioning  the global depth $D(\bmy|\bmX)$  on this neighborhood yields its \emph{$\beta$-localized depth},
\begin{equation} \label{localVanBever}
D_{\rm{\beta}}(\bmy|\bmX)=  D(\bmy| P_{\beta, \bmy,\bmX})\,,
\end{equation}
where $P_{\beta, \bmy,\bmX}$ is the conditional distribution of $\bmX$, conditioned on $D_{\rm{\beta}} (P_{\bmy,\bmX})$.
If $D$ is affine invariant, so is its $\beta$-localized depth. Obviously, if $\beta=1$, the global depth is obtained. If $\beta \to 0$, the localized depth $D_{\rm{\beta}}(\bmy|\bmX)$ does \emph{not} converge to the density of $\bmX$ but rather to a constant which is positive and reflects local asymmetry  for $\bmy$ within the support of $\bmX$, and which (usually) vanishes outside.



{Figure~\ref{fig:locals} exhibits regions of the localized Mahalanobis, zonoid and halfspace depths using the localization approach of \cite{PaindaveineVB13}. The three samples contain $300$ observations from a moon-shaped, a bimodal, and a trimodal distribution, respectively. The localization parameter is $0.33$.
It is seen that localization is able to improve the fit of a multimodal distribution. On the other hand, spurious high-depth zones can arise (here, one for the bimodal and four for the trimodal data), due to the centrality-proneness of depth. In applications the localization parameter has to be properly chosen, depending on the data and the problem at hand. This requires prior information or (e.g. in supervised classification) tuning.


\begin{figure}[!h]
	\begin{center}
	\begin{tabular}{cccc}
		& \,\,Moon-shaped & Bimodal & Trimodal \\
		\begin{sideways}\,\,\,\quad Mahalanobis depth\end{sideways} & \includegraphics[keepaspectratio=true,width = 0.285\textwidth, trim = 10mm 10mm 10mm 20mm, clip, page = 1]{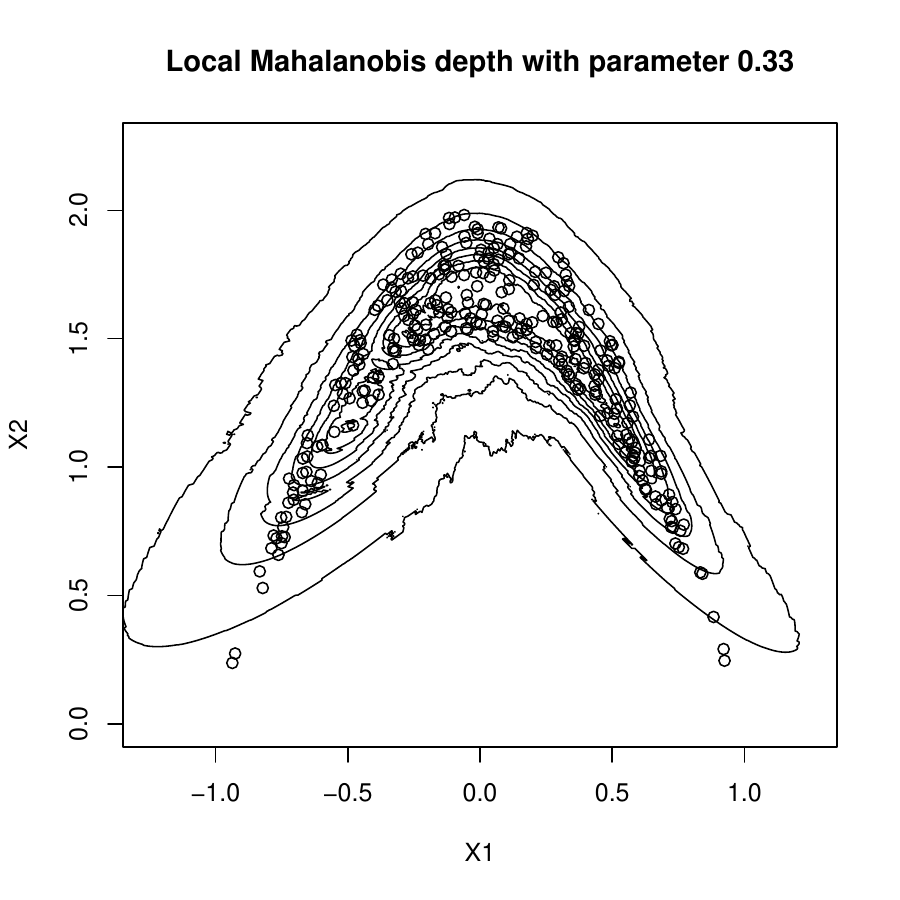} & \includegraphics[keepaspectratio=true,width = 0.285\textwidth, trim = 10mm 10mm 10mm 20mm, clip, page = 1]{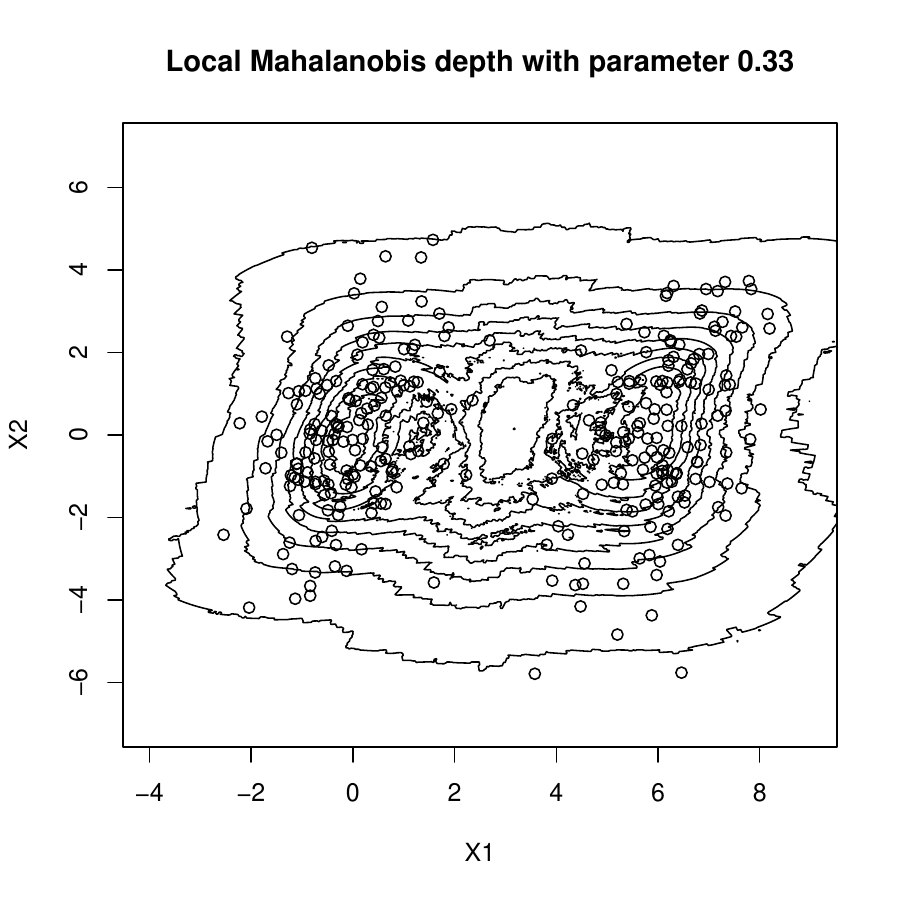} & \includegraphics[keepaspectratio=true,width = 0.285\textwidth, trim = 10mm 10mm 10mm 20mm, clip, page = 1]{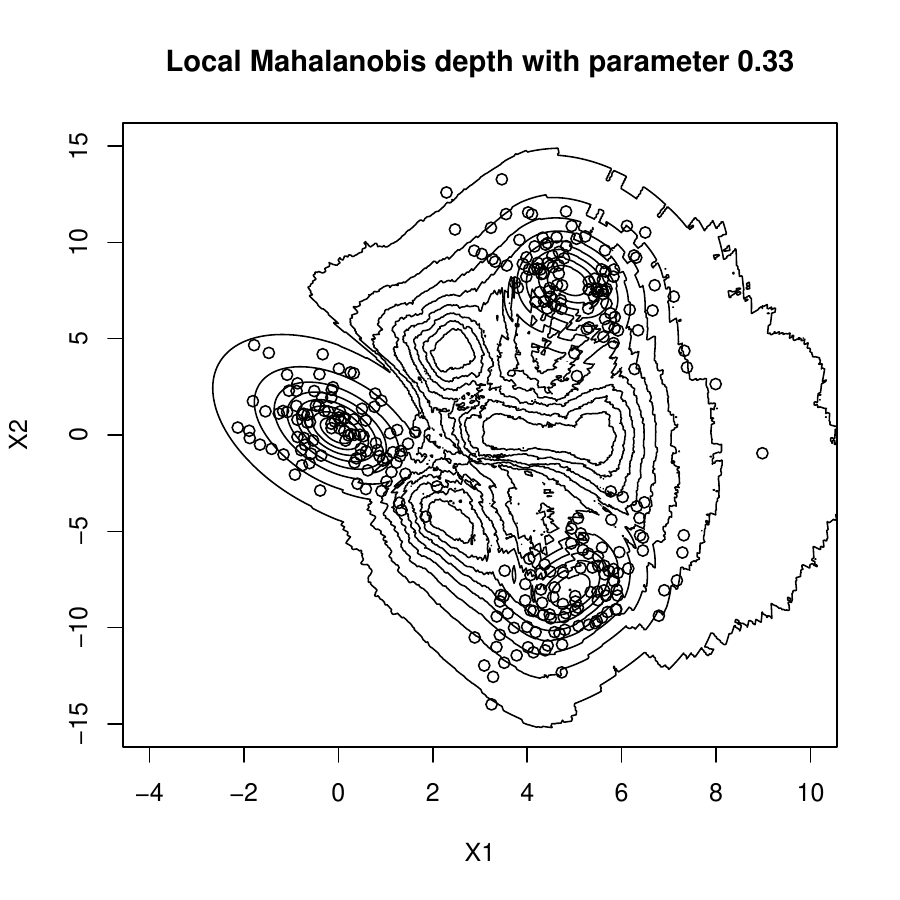} \\
		\begin{sideways}\quad\qquad zonoid depth\end{sideways} & \includegraphics[keepaspectratio=true,width = 0.285\textwidth, trim = 10mm 10mm 10mm 20mm, clip, page = 2]{pic-local-depths-X1.pdf} & \includegraphics[keepaspectratio=true,width = 0.285\textwidth, trim = 10mm 10mm 10mm 20mm, clip, page = 2]{pic-local-depths-X2.pdf} & \includegraphics[keepaspectratio=true,width = 0.285\textwidth, trim = 10mm 10mm 10mm 20mm, clip, page = 2]{pic-local-depths-X3.pdf} \\
		\begin{sideways}\,\,\,\qquad halfspace depth\end{sideways} & \includegraphics[keepaspectratio=true,width = 0.285\textwidth, trim = 10mm 10mm 10mm 20mm, clip, page = 3]{pic-local-depths-X1.pdf} & \includegraphics[keepaspectratio=true,width = 0.285\textwidth, trim = 10mm 10mm 10mm 20mm, clip, page = 3]{pic-local-depths-X2.pdf} & \includegraphics[keepaspectratio=true,width = 0.285\textwidth, trim = 10mm 10mm 10mm 20mm, clip, page = 3]{pic-local-depths-X3.pdf} \\
	\end{tabular}
	\end{center}
	\caption{Localized depth regions for $300$ points from moon-shaped (left), bimodal (middle), and trimodal (right) using localized~\citep{PaindaveineVB13} Mahalanobis (top), zonoid (middle), and halfspace (bottom) depths with localization parameter~$0.33$.}\label{fig:locals}
\end{figure}
}

\subsection{Monge-Kantorovich depth}
{A global depth that allows for non-convex level sets but retains nestedness and a `common center' is proposed in \cite{ChernozhukovGHH17} and mentioned as Monge-Kantorovich (MK) depth. It is based on the theory of translocation of masses. Let $\bmX$ and $\bmY$ be random vectors in $\IR^d$ having finite second moments. Then there exists a function $g_{\bmX,\bmY}$ with $\bmY = g_{\bmX,\bmY}(\bmX)$, called MK translocation, that transports the probability mass of $\bmX$ to that of $\bmY$ at
minimum quadratic cost. If in addition $\bmX$ and $\bmY$ possess densities, $g_{\bmX,\bmY}$ is (almost surely) unique and has an inverse $g_{\bmY,\bmX}$. }

{ Recall that, if $\bmX$ has an elliptically symmetric distribution, the level sets of any affine invariant depth coincide with the density level sets.
This is the starting point and benchmark of the MK depth.
A general non-elliptically distributed $\bmX$ is transformed to a spherically distributed one, \emph{viz.} to a random vector $\bmU$ that has uniform distribution on the unit ball $B^{d-1}=\{\bmx \in \IR^d : ||\bmx|| \le 1\}$.
If $\bmX$ has finite second moments, the transformation is done by an MK translocation of masses $g_{\bmX,\bmU}$, which transports the probability mass of the given distribution to the uniform distribution  at minimal quadratic cost. This transform serves as a \emph{center-outward distribution function} of $\bmX$, and its inverse $g_{\bmU,\bmX}$ as a \emph{quantile  function}. For further details and a more general setting, see \cite{HallindelBCAM20}.}

{The Tukey central regions of $\bmU$ are concentric spheres. Their images by $g_{\bmU,\bmX}$ are mentioned as the \emph{MK central regions} of $\bmX$, which form the level sets of the \emph{MK depth}.
From their construction follows that, if $\bmX$ has an elliptical distribution, MK regions and MK depth coincide with Tukey regions and Tukey depth;  further, that in case of a centrally or angularly symmetric distribution of $\bmX$ the MK depth takes its maximum at the center of symmetry.}

{The MK depth is invariant to translation, rotation and global rescaling, but generally not affine invariant. It reflects asymmetries of the distribution. 
The MK depth is consistently estimated by its sampling version.
MK central regions refer to kind of a `common center' and follow more closely the shape of a given distribution than convex or starshaped regions can do.
They give rise to distribution free notions of multivariate quantiles, ranks, and signs \citep{HallindelBCAM20}.

Given two random vectors $\bmX$ and $\bmZ$ having densities and finite second moments, we may consider the mapping $g_{\bmU,\bmZ} \circ g_{\bmX,\bmU}$. This mapping transforms any MK central region of $\bmX$ that has a certain probability mass to the MK region of $\bmZ$ having the same probability mass.}

\subsection{Functional depths}\label{sec6.2}

So far, the discussion has been restricted to depths in $d$-space. However, notions of depth for functional data
have gained much interest in the past decade.
Consider a space $E$ of functions $[0,1]\to \IR^m$ equipped with the supremum norm and $E'$ its dual space of continuous linear functions $E\to \mathbb{R}^m$.
A functional data depth is a real-valued functional that measures how deep a function $\bmy\in E$ is located in a given finite cloud
$\bmX=\{\bmx^1, \dots, \bmx^n\}$ of functions $\in E$.
{Several multivariate depths extend verbatim to a functional setting. E.g., the lens depth, being based on distances between points, is well-defined on general metric spaces; see \cite{CholaquidisFGM20}, who include an application of lens depth to classification of data in Riemannian manifolds.
Other depths, like the Tukey depth, though being formally defined for functional data, collapse to zero in this setting \citep{DuttaGC11}.
For a recent survey on notions of functional depth and their properties, see \cite{GijbelsN17}.}

Most known functional depths belong to two types, which build on multivariate depths like those discussed above. Either they are of \emph{integral type} \citep{NagyGOH16},
\begin{equation}\label{defintFD}
 D(\bmy|\bmX)= \int_0^1 D^m(\bmy(t)|\bmX(t))dt\,,
 \end{equation}
or of \emph{infimum type} \citep{MoslerP12}
\begin{equation}\label{definfFD}
 D(\bmz|\bmX)= \inf_{\varphi\in \Phi} D^m(\varphi(\bmy|\varphi(\bmX))\,,
 \end{equation}
where $D^m$ is an $m$-variate data depth,  $\Phi$ is a proper subset of linear functionals in ${E'}^d$, and $\varphi(\bmX)$ is the transformed data cloud $\{\varphi(\bmx^1), \dots, \varphi(\bmx^n)\}$.  Population versions are similarly defined.

A depth of integral type (\ref{defintFD}) is just an average  of multivariate depth values attained at all `times' $t$.
Note that in definition (\ref{definfFD}) of infimum-type depth each $\varphi$ may be interpreted as a particular aspect of $\bmy$
that is of interest and which is represented in $m$-dimensional space. A
depth of infimum type (\ref{definfFD}) is given as the smallest multivariate depth of $\bmy$ under all these aspects.

It is obvious from definitions (\ref{defintFD}) and (\ref{definfFD}) that the properties of these functional depths depend essentially on the properties of the involved multivariate depth. For a comprehensive treatment, the reader is referred to \cite{GijbelsN17}; {see also \cite{NietoRB16} and \cite{KuelbsZ13}.} \cite{ChowdhuryC19} consider spatial depth and apply it to functional quantile regression.

{\subsection{Depth of a fit}}
{
\cite{RousseeuwH99} propose another view on depth as follows.
\begin{itemize}
\item \textbf{Nonfit:} A depth $D(\theta|\bmx_1,\dots,\bmx_n)$ is defined as the smallest proportion $k/n$ of data points that need to be removed to make $\theta$ a nonfit.
\item \textbf{Loc:} Given data $\bmx_1,\dots,\bmx_k\in \IR^d$, the point $\theta\in \IR^d$ is called a \emph{nonfit} if $\theta$ lies outside the convex hull $\conv\{\bmx_1,\dots,\bmx_k\}$.
\end{itemize}
They demonstrate that \textbf{Nonfit} and \textbf{Loc} yield the Tukey depth $D_H(\theta|\bmx_1,\dots,\bmx_n)$.
The idea of reducing a given data set until a certain criterion is satisfied is not restricted to locational problems.
In their ingenious paper \cite{RousseeuwH99} transfer it to linear regression,
\[ \bmy = [\boldmath{1}, \bmX] \theta\,,\]
where $\bmy\in\IR^d$ and an $n\times d$ matrix $\bmX=[\bmx_1, \dots, \bmx_n]^T$ are given, $\boldmath{1}$ is a column of ones, and $\theta\in \IR^{d+1}$ a parameter vector to be determined.
They replace the criterion \textbf{Loc} by
\begin{itemize}
\item \textbf{Regr:} $\theta$ is a nonfit to the linear regression problem if there exist $\bmv\in \IR^d$ and $\gamma\in \IR$
so that
\begin{equation}\label{regressioncriterion}
(\bmv'\bmx_i- \gamma) r_i >0 \quad \text{for all} \;\; i \,,
\end{equation}
where $\bmr = (r_1, \dots, r_n)^T = \bmy- [\boldmath{1}, \bmX] \theta$ is the vector of residuals.
\end{itemize}
Condition (\ref{regressioncriterion}) says that there exists a hyperplane
$H=\{\bmx\in \IR^d : \bmv'\bmx=\gamma\}$  separating the $\bmx$-data so that $r_i>0$ holds for all $\bmx_i$ on one side of $H$, while $r_i<0$ holds for the remaining $\bmx_i$. \textbf{Nonfit} and \textbf{Regr} define the depth of a vector $\theta$ of candidate regression parameters. \cite{RousseeuwH99} call it  \emph{regression depth} and solve the regression problem by searching a vector $\theta_{max}$ of maximal regression depth.

Clearly, a candidate $\theta$ that yields only residuals $r_i$ of the same sign is a nonfit and has regression depth null. In general, a nonfit $\theta$ can be always improved by another $\theta$ that changes at least one residual sign. Regression depth is the minimum proportion of residual signs to be changed to make $\theta$ a nonfit. In this sense, $\theta_{max}$ provides a regression that is `most central' in the data. Regression depth is affine invariant, and deepest regression is robust having breakdown value $1/3$. For details see \cite{RousseeuwH99}, who provide many more properties including a population version of their approach. They also propose a regression version of simplicial depth, which has convenient asymptotics.

Other multivariate depths, among them projection depth, have been transferred to the linear regression context, too; see the general treatment and survey by \cite{Zuo21}.

The depth-of-a-fit approach combines a general nonparametric principle of robustness (\textbf{Nonfit}) with a criterion of `fitness' of a parameter, which hypothesizes a specific semiparametric model.
\cite{Mizera02} develops the approach beyond linear regression to what he calls \emph{tangent depth}, considering differentiable $\theta$'s that live on a manifold and are evaluated by a differentiable criterion function. Those criterion functions can e.g., as with regression depth, be based on residuals or, as it is done with \emph{location-scale depth} \citep{MizeraM04}, on a likelihood function. However, for more recent developments, since this line of research is beyond the scope of the present article, the reader is referred to \cite{KustoszMW16} and the literature referenced there.}

\section{Concluding remarks}\label{sec7}

Several popular notions of multivariate depth functions have been considered and compared, with a view to practical applications.
Different depths yield different central ( = trimmed) regions and different medians.
While many notions are affine invariant and, thus independent of a coordinate system, some are only rigid-body invariant (regarding translation and orthogonal transform), but can be made affine invariant through whitening the data. The depth notions differ in their analytical properties, particularly in the information they carry about the underlying distribution and its center. E.g., the zonoid depth characterizes the whole distribution, while the Mahalanobis depth determines the first two moments only.  Also, for numerical calculations, continuity is an issue.
Some notions (like halfspace and simplicial depth) are robust against extremely outlying data, others are not.
{If only ordinal information about distances is available, lens depth may be used, see \cite{KleindessnerVL17}.}
These and several other properties {may} guide the choice of a proper depth notion in a specific application.

Moreover, as all depth notions (besides moment Mahalanobis depth) are more or less computationally intensive, computational feasibility is a key aspect in this choice. For all notions considered here, exact and/or approximate algorithms exist, which are implemented in \texttt{R}-packages like \texttt{ddalpha}. But computational complexity of these procedures ranges from
$O=(n)$ to $O(n^{d+1})$. These complexities have been presented above together with calculation times of exact procedures for moderate $n$ and $d$.


For some depths on higher-dimensional data and larger sample sizes, approximate algorithms have to be employed. Their complexities are given above as well. Regarding the accuracy of approximate procedures, specifically the random Tukey depth has been numerically compared with the exact halfspace ( = Tukey) depth, using a fixed number of random directions.
General strategies for large and high-dimensioned data have been discussed, too.

Many of these remarks apply also to depth statistics for functional data, as the functional  depths usually build on multivariate depth notions and operate on discretized versions of the data.

We close with a few rough conclusions regarding the practice of data analysis.
\begin{enumerate}
\item If the data is asymmetric, centrality should be measured by a depth that reflects the data's shape.
\item A depth should be chosen that is at least invariant to shifts and rotations and that is efficiently computable for the given dimension and size of the data.
\item For large data sets, spatial, $\IL_2$, and Mahalanobis depths are efficiently and exactly calculated. Next to them, lens and zonoid depths.
\item For the other depths, approximate procedures are available. However, the accuracy of e.g.\ the random Tukey depth declines rapidly when dimension increases.
\item Sphering can be costly in terms of precision. If affine invariance is not needed, it should be avoided.
\item There is a trade-off between robustness and computational complexity. If the data appears to be contaminated, robustified Mahalanobis depth may be employed in case of elliptically symmetric data, and spatial or halfspace depth otherwise. If not, zonoid depth is a good choice.
\item  In case of missing values, zonoid depth can be used  with mean imputation.
\end{enumerate}

We have covered {a few widely used notions of depth to measure outlyingness of a point in $\IR^d$}. Many more have been and are still being proposed in the literature. To be meaningful they should be
sufficiently invariant, reflect asymmetries of the data, and be computationally feasible for practically relevant dimensions $d$ and data sizes $n$, with either exact or sufficiently precise approximative procedures.

\section*{Acknowledgments}
We thank Stanislav Nagy for many useful remarks on a previous version of this paper, and Yijun Zuo for valuable hints to the literature. {Also the remarks of two anonymous referees are gratefully acknowledged.}



\end{document}